\documentclass{aa}  
\usepackage{natbib}
\usepackage{graphicx}
\usepackage{txfonts}
\usepackage{xcolor}
\usepackage{soul}

\usepackage{ctable} 

\begin{document}

   \title{Validation of EUHFORIA cone and spheromak Coronal Mass Ejection Models}
   \authorrunning{L. Rodriguez et al.}
   \titlerunning{Validation of EUHFORIA}


   \author{L. Rodriguez
          \inst{1},
          D. Shukhobodskaia\inst{1}, A. Niemela \inst{1,}\inst{2}, A. Maharana\inst{1,}\inst{2}, E.Samara \inst{3}, C. Verbeke \inst{2},
          J. Magdalenic\inst{1,}\inst{2}, R. Vansintjan \inst{1}, M. Mierla \inst{1,} \inst{4}, C. Scolini \inst{1}, R. Sarkar \inst{5}, E. Kilpua \inst{5}, E. Asvestari \inst{5},  K. Herbst \inst{6},  G. Lapenta \inst{2}, A.D. Chaduteau \inst{7},   J. Pomoell \inst{5}, \and S. Poedts \inst{2,} \inst{8}
          }

   \institute{Solar-Terrestrial Centre of Excellence – SIDC, Royal Observatory of Belgium; Avenue Circulaire 3, 1180 Brussels, Belgium\\
              \email{luciano.rodriguez@observatory.be}
         \and
             CmPA/Department of Mathematics, KU Leuven, Celestijnenlaan 200 B, 3001 Leuven, Belgium 
         \and NASA Goddard Space Flight Center, Greenbelt, MD, USA
         \and Institute of Geodynamics of the Romanian Academy, Bucharest, Romania
         \and Department of Physics, University of Helsinki, PO Box 64, 00014, Helsinki, Finland
         \and Institut f\"{u}r Experimentelle und Angewandte Physik, Christian-Albrechts-Universit\"{a}t zu Kiel, D-24118 Kiel, Germany
         \and Blackett Laboratory, Imperial College London, London SW7 2AZ, United Kingdom
         \and Institute of Physics, University of Maria Curie-Sk{\l}odowska, Lublin, Poland
         \\
             }

   \date{Received February 20, 2024}

 
  \abstract
   {}
   {We present the validation results for arrival times and geomagnetic impact of Coronal Mass Ejections (CMEs), using the cone and spheromak CME models implemented in EUropean Heliospheric FORecasting Information Asset (EUHFORIA). Validating numerical models is crucial in ensuring their accuracy and performance with respect to real data.}
   {We compare CME plasma and magnetic field signatures, measured in situ by satellites at the L1 point, with the simulation output of EUHFORIA. The validation of this model was carried out by using two datasets in order to ensure a comprehensive evaluation. The first dataset focuses on 16 CMEs that arrived at the Earth, offering specific insights into the model's accuracy in predicting arrival time and geomagnetic impact. Meanwhile, the second dataset encompasses all CMEs observed over eight months within Solar Cycle 24, regardless of whether they arrived at Earth, covering periods of both solar minimum and maximum activity. This second dataset enables a more comprehensive evaluation of the model's predictive precision in term of CME arrivals and misses.}
   {Our results show that EUHFORIA provides good estimates in terms of arrival times, with root mean square errors (RMSE) values of 9 hours. Regarding the number of correctly predicted ICME arrivals and misses, we find a 75\% probability of detection in a 12~hours time window and 100\% probability of detection in a 24~hours time window. The geomagnetic impact forecasts, measured by the $K_p$ index, provide different degrees of accuracy, ranging from 31\% to 69\%. These results validate the use of cone and spheromak CMEs for real-time space weather forecasting.
}
   {}


   \maketitle
%

\section{Introduction} \label{S-Introduction}

   Coronal Mass ejections (CMEs) are remarkable transient events that affect the solar system. They involve substantial release of both magnetic field and plasma from the Sun's corona, typically travelling with speeds ranging from 400 to 1000~km~s$^{-1}$, and occasionally exceeding 2000~km~s$^{-1}$ \citep{Hundhausen1994,Dryer2012,Liou2014}.
Depending on their initial speed and that of the ambient solar wind, the majority of CMEs typically arrive to the orbital distance of the Earth within 1--4 days. Upon reaching our planet, CMEs can induce geomagnetic storms through interaction with the Earth's magnetosphere. The severity of these storms depends on the internal magnetic field configuration and plasma properties of the CME \citep[e.g.][]{Schwenn2006, Temmer2021, Kilpua2017SSR,Koskinen2017}. 
CMEs detected by a spacecraft in-situ are also called interplanetary CMEs (ICMEs) \citep[e.g.][]{Rodriguez2011,Kilpua2017LRSP}. ICMEs are characterised by specific signatures observed in the magnetic field and plasma data \citep[][]{Zurbuchen2006}.

Halo CMEs refer to a specific type of CMEs observed in white-light coronagraph images,  where the expanding structure appears as a \emph{halo} around the occulting disk, that occurs when a CME travels either towards or away from the observer. 
When observed from Earth, these eruptive events are crucial for space weather: their detection (for front-sided events) indicates the potential propagation of the eruption along or in close proximity of the Sun--Earth line  \citep{Howard1982,Schwenn2006,Rodriguez2009halos}. 

Historically, detection of halo CMEs along the Sun-Earth line relied heavily on instruments such as Large Angle and Spectroscopic COronagraph \citep[LASCO;][]{brueckner1995}  aboard the  SOlar and Heliospheric Observatory \citep[SOHO;][]{Domingo1995}, positioned at the Lagrange point L1 of the Sun--Earth system. 
Additional insights into associated signatures in the low corona, such as eruptive filaments, coronal dimmings, ``EIT waves'' and post-eruption arcades, were gained through Extreme-ultraviolet (EUV) observations \citep[e.g.][]{Zhukov2007}, notably from instruments like the Extreme-ultraviolet Imaging Telescope \citep[EIT;][]{Delaboudiniere95}.

However, the detection of CMEs improved with the launch of the Solar-TErrestial RElations Observatory \citep[STEREO;][]{Kaiser2008}. The twin STEREO spacecraft provided a view from a location away from the Sun-Earth line by means of the COR coronagraphs and the Extreme Ultraviolet Imagers (EUVI) of the Sun Earth Connection Coronal and Heliospheric Investigation instrument suites \citep[SECCHI;][]{Howard2008}. When the STEREO spacecraft are positioned away from the Sun--Earth line they enable tracking of Earth-directed CMEs based on side view observations \citep[e.g.][]{Davies2009,Moestl_2011,Rodriguez2020}. This provides a  very important viewpoint needed for an early characterization of Earth-directed CMEs. In particular for determining the CME propagation direction, speed and acceleration with minimal projection effects, therefore allowing a more accurate estimation of CME arrival times at Earth. This information, together with the knowledge of the internal magnetic field configuration of the CME, are of crucial importance for space weather.  

The EUropean Heliospheric FORecasting Information Asset \citep[EUHFORIA,][]{Pomoell2018} is a space weather forecasting-targeted inner heliosphere physics-based model. It consists of two coupled domains, the coronal (which focuses on processes below 0.1~au) and the heliospheric (which focuses on the heliosphere starting at 0.1~au). The first one uses synoptic magnetograms in order to compute magnetic field and plasma parameters at 0.1~au using an adaptation of the Wang-Sheeley-Arge empirical model \citep[WSA,][]{2003Arge}. The second domain employs a model that takes as input the output at 0.1~au and solves the three-dimensional (3D) time-dependent ideal Magnetohydrodynamic (MHD) equations in the HEEQ system, at a prescribed resolution. Finally, CMEs can then be incorporated into the heliospheric simulation of EUHFORIA using different CME models, such as the cone model \citep{Xie2004} which does not prescribe an internal magnetic field configuration for the CME, and the more complex spheromak \citep[][]{Verbeke2019,Scolini2019} flux rope CME model, which has a prescribed internal magnetic field configuration.

During the European Union Horizon 2020 project ``EUHFORIA 2.0'',  we carried out a validation of the cone and spheromak CME models, which is described in this work. We evaluated the models both for the CME arrival time accuracy and for the possibility of predicting their geomagnetic impact, based on the internal magnetic field profile and plasma parameters.

The paper is organized as follows. Section \ref{Section_Event_Selection} presents the selection of events for the two datasets that were used in the validation, and the collection of the necessary input data. Section \ref{section_overview} provides a brief overview of the models and the metrics used in the study. Then, in Section \ref{section_resultsA} we compare the results for arrival times and ICME internal characteristics, including geoeffectiveness, for the first dataset. In Section \ref{section_resultsB}, using the second dataset, we evaluate how accurate (in terms of hit and miss) CME arrival predictions in EUHFORIA are. Finally, in Section \ref{Summary} we summarize our results and provide conclusions.


\section{Selection of events and collection of input data}
\label{Section_Event_Selection}
In this section, we outline our approach to event selection and data gathering essential for the study. The validation was performed using two distinct datasets: one containing 16 CMEs that arrived at the Earth (Event list A), another comprising all CMEs observed over eight months within Solar Cycle 24, regardless of their Earth impact (Event list B). The reasoning for having two datasets is to study and validate different aspects. With the first dataset, we test how the CMEs simulated by EUHFORIA compare to in situ satellite data collected at the L1 point for the corresponding ICMEs. The second dataset allows us to evaluate the accuracy of EUHFORIA in predicting CME arrivals, based on a scenario that more closely resembles that of a forecasting situation in real time, where all CMEs are considered.

\subsection{Events list}

\subsubsection{Event list A - Selected events}
\label{subsection_events}
This section describes the list of events (CME--ICME pairs) that are used for the first part of study. The CMEs were chosen when coronagraphic observations from at least two out of three spacecraft (STEREO-A, STEREO-B, SOHO) were available, so that 3D-reconstructed geometric parameters could be used to drive EUHFORIA \citep{Pomoell2018,Scolini2019}. This requirement sets the start range of our possible CME candidates in 2007 (start of STEREO data availability). 
Furthermore, we required good availability of EUV and magnetic field data of the source region, which needs to be visible on-disk from the Earth's point of view, to ensure a clear characterization of the CME source.
We excluded events with interacting CMEs, to reduce the complexities that could influence the analysis. Finally, the corresponding ICMEs were required to arrive to the Earth, so that  simulation results could be compared with in situ data at the L1 point. We created a list of 16 events, shown in Table \ref{table_eventslist}. This list is non-exhaustive, events were selected by visually inspecting the data and choosing clear cases for which the above mentioned conditions were fulfilled.


\begin{table}[ht]
\caption{The table includes a list of CME--ICME pairs used in this study (Column 2 and Column 3), with the provided times corresponding to the first image in LASCO-C2 for the CMEs, and the start of the disturbance in OMNI data for the ICMEs. Additionally, it presents the date and time of the GONG magnetogram (Column 4) used for EUHFORIA runs, along with simulated ICME arrival time for cone (Column 5) and spheromak (Column 6) models. The provided times are in UTC format.}

\label{table_eventslist}
\resizebox{\hsize}{!}{
\begin{tabular}{c|c|c|c|c|c}
\hline
Event & CME & ICME & Magnetogram & Cone  & Spheromak           \\
No. & & & & model  & model  \\
\specialrule{.15em}{.09em}{.09em} 
\multicolumn{6}{c}{2010} \\ 
\specialrule{.15em}{.09em}{.09em} 
1         & 03/04 10:33 & 05/04 08:26  & 28/03 06:00 & 05/04 16:53 & 06/04 01:16  \\
\specialrule{.15em}{.09em}{.09em} 
\multicolumn{6}{c}{2011}
\\ 
\specialrule{.15em}{.09em}{.09em} 
2         & 06/09 23:05 & 09/09 12:42  & 04/09 12:00 & 09/09 20:58 & 10/09 06:04
\\
\hline
3         & 22/09 10:48 & 25/09 08:44 & 24/09 12:00 & 24/09 08:44 & 25/09 17:19 \\
\hline
4         & 24/09 12:48 & 26/09 12:34 & 24/09 12:00 & 26/09 04:49 & 26/09 14:34 \\

\specialrule{.15em}{.09em}{.09em} 
\multicolumn{6}{c}{2012} \\ 
\specialrule{.15em}{.09em}{.09em} 
5         & 07/03 00:24 & 08/03 11:03  & 06/03 23:00 & 08/03 01:09 & 08/03 01:49  \\
\hline
6         & 13/03 17:36 & 15/03 13:06  & 07/03 12:00 & 15/03 14:53 & 15/03 19:04 \\
\hline
7         & 12/05 00:00 & 15/05 02:00 & 10/05 23:30 & 14/05 16:38 & 15/05 04:18 \\
\hline
8         & 12/07 16:48 & 14/07 18:09 & 09/07 23:30 & 14/07 19:48 & 14/07 11:53 \\
\hline
9         & 28/09 00:12 & 30/09 23:05 & 23/09 06:00 & 30/09 02:58 & 30/09 09:53 \\
\specialrule{.15em}{.09em}{.09em} 
\multicolumn{6}{c}{2013}  \\ 
\specialrule{.15em}{.09em}{.09em} 
10        & 15/03 07:12 & 17/03 05:59 & 13/03 12:00 & 17/03 04:18 & 17/03 01:53 \\
\hline
11        & 11/04 07:24 & 13/04 22:54 & 08/04 06:00 & 13/04 20:14 & 13/04 20:19 \\
\hline
12        & 29/09 22:12 & 02/10 01:54  & 28/09 18:00 & 02/10 15:34 & 02/10 14:19\\
\specialrule{.15em}{.09em}{.09em} 
\multicolumn{6}{c}{2014} \\ 
\specialrule{.15em}{.09em}{.09em} 
13        & 07/01 18:24 & 09/01 19:40  & 04/01 18:00 & 09/01 15:28 & 10/01 14:28\\
\hline
14        & 10/09 18:00 & 12/09 15:53  & 06/09 18:00 & 12/09 21:53 & 12/09 23:00 \\
\specialrule{.15em}{.09em}{.09em} 
\multicolumn{6}{c}{2017}  \\ 
\specialrule{.15em}{.09em}{.09em} 
15        & 04/09 22:36 & 06/09 23:47 & 04/09 18:00 & 06/09 13:32 & 07/09 01:14 \\
\hline
16        & 06/09 14:13 & 07/09 23:02 & 04/09 18:00 & 08/09 01:28 & 08/09 05:23\\
\hline
\end{tabular}}
\end{table}

\subsubsection{Event list B - CMEs over eight months}
\label{subsubsection_eventsB}

To assess EUHFORIA performance in predicting the arrival or non-arrival of the targeted CMEs,
we utilize a second dataset.
In this second dataset, we do not require an ICME counterpart at the Earth for the observed CMEs.  Rather, we used all the CMEs observed during eight full months. Four months were taken during solar minimum (June - September 2010) and the other four during solar maximum (June - September 2012). For the former period (2010), we found originally 357 events, taken from the SOHO/LASCO CME catalog \citep{Gopalswamy2009} \footnote{https://cdaw.gsfc.nasa.gov/CME\_list/}. After excluding events with data gaps, visible only by one spacecraft, faint events with uncollectable input parameters and implementing thresholds for velocity (we selected only CMEs faster than 350 km~s$^{-1}$) and angular width (only CMEs wider than 60$^{\circ}$ were considered), the list was reduced to 24 events. From those cases, 12 are frontsided and were simulated with EUHFORIA. Since we are considering the ability of EUHFORIA to predict the arrival or miss at Earth's position, we do not include backsided events, which are obviously not going to arrive to the Earth. For the period in 2012 (solar maximum), originally 857 events were collected, after removing events with gaps and taking same velocity and angular width constraints as during solar minimum, 191 events remained. Out of those, 36 frontsided events were simulated with EUHFORIA.

\subsection{Collection of input data}
\label{subsection_input}
The knowledge of realistic CME intrinsic parameters is crucial for space weather forecasting using first principle heliospheric models and semi-empirical CME models \citep[e.g.,][and references therein]{Kilpua2019a}. Regarding EUHFORIA, the interest is to constrain the geometry and kinematics for CMEs using the cone and spheromak CME models. Furthermore, for the latter one also needs to determine magnetic field parameters that will serve as input to the model.
Below we outline, how input parameters were collected for Event list A, comprising selected Events. For the Event list B, covering CMEs observed over eight months, only cone CME runs were utilized. Magnetic field configuration was not needed for this part of the study, which primarily aims at assessing model performance in forecasting CME arrivals and misses. Input parameters for these EUHFORIA simulations were gathered using the StereoCat tool \footnote{https://ccmc.gsfc.nasa.gov/analysis/stereo/}, which is better suited for data analysis in a real time forecasting environment.

\subsubsection{Geometric and kinematic reconstruction of CMEs}
\label{subsubsection_geoparam}

In this study for the Event list A - Selected events, we use the Graduated Cylindrical Shell (GCS) model \citep{Thernisien2009,Thernisien2011} to determine the 3D morphology, and the propagation direction of CMEs. 
It consists of a tubular section forming the main body of the structure attached to two cones that correspond to the ``legs'' of the CME. As the model consists of a single geometric surface, it does not provide a description of the internal structure of the CME. 

The model fits the geometrical structure of the CME as observed by white-light coronagraphs such as SOHO/LASCO and STEREO/COR2. The fitting can be performed from single or multiple spacecraft data. The results are however more reliable when multiple and well-separated vantage points are used \citep[e.g.][]{Rodriguez2011}. The following parameters can be obtained:
\begin{itemize}
    \item propagation longitude ($\phi$)
    \item propagation latitude ($\theta$)
    \item  half-angular width ($\alpha$)
    \item aspect ratio ($\kappa$), that is the rate of expansion vs the height of the CME
    \item tilt angle ($\gamma$) with respect to the solar equator 
    \item leading-edge height ($h$) of the CME
\end{itemize}

\begin{figure}
  \centering
  \resizebox{\hsize}{!}{\includegraphics{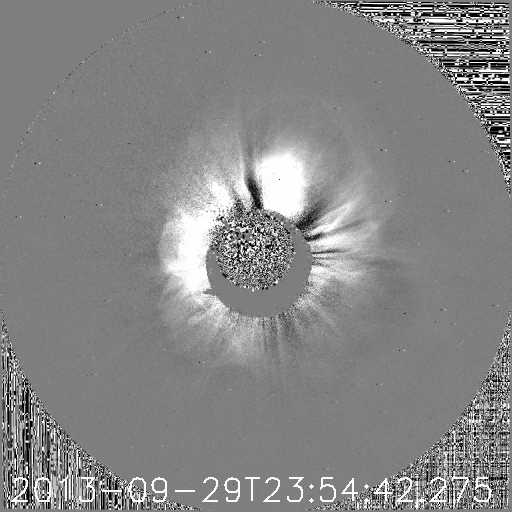}
  \includegraphics{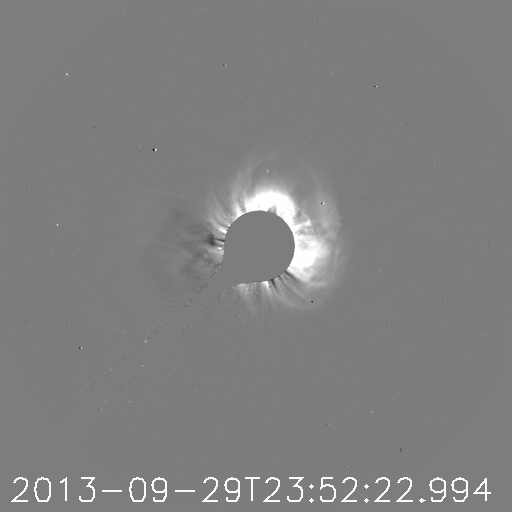}
  \includegraphics{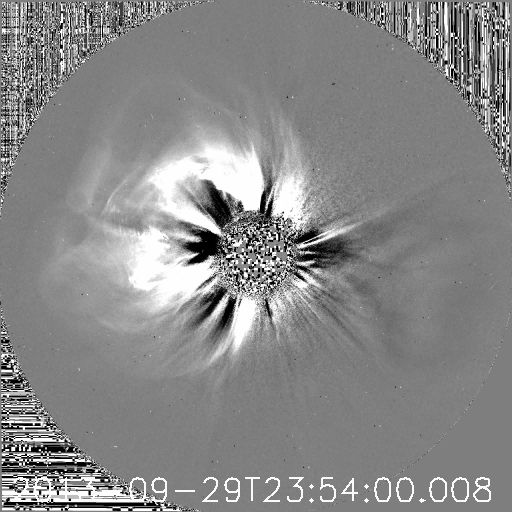}}
  \resizebox{\hsize}{!}{
  \includegraphics{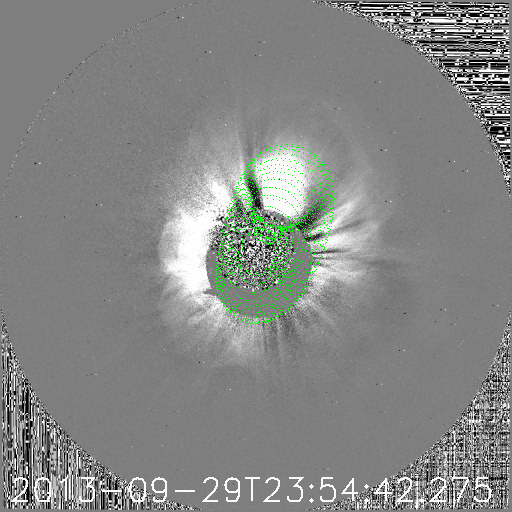}
  \includegraphics{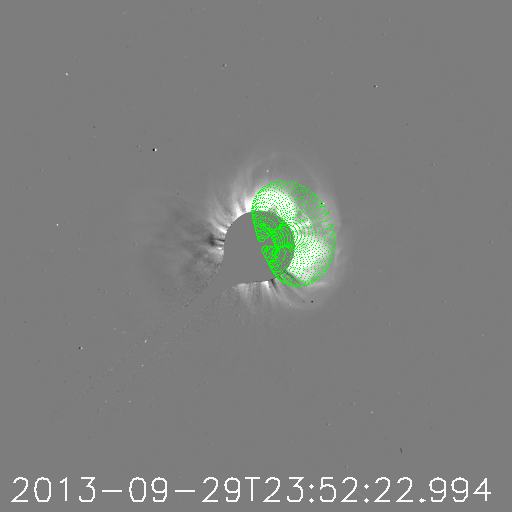}
  \includegraphics{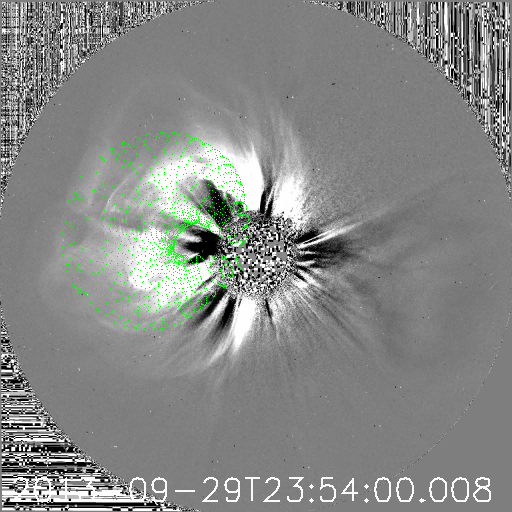}}
  \caption{
    GCS reconstruction (bottom row) of the CME observed (top row) by COR2-B (left panel), LASCO-C3 (middle panel), and by COR2-A (right panel) at 23:54 UT on 29 September 
    2013.
  }
  \label{Fig:GCSreconstr}
\end{figure}

The fitting was performed for different consecutive moments of time, for each CME, in order to derive true 3D velocity vectors of the expanding structure. An example of a fit for the CME observed on 29 September 2013 is shown in Figure~\ref{Fig:GCSreconstr}. The GCS fitting is shown as the green wireframe overlaid on the coronagraph images at each of the three spacecraft. 

Note that due to the symmetry of the model, the fit is not perfect in the case of complex CMEs. \cite{Thernisien2006} mentioned several sources of errors in the model, among them the errors intrinsic to the fitting method, and the errors arising from the subjectivity of the observer. Regarding the intrinsic errors, it is important to note that different parameters may fit the same CME well (i.e., the solution may not be unique). This is especially true when fitting single-spacecraft data, which provide a single 2D projection of a 3D structure.
\cite{Thernisien2009} have estimated the errors for the GCS model using a sensitivity analysis method. For the 26 events studied they found a mean value of $\pm~4.3^\circ$ (with a maximum value of $16.6^\circ$) in longitude, and $\pm~1.8^\circ$ degrees (max value of $3.7^\circ$) in latitude. 
\cite{Verbeke2023} estimated that the largest errors are found when only one viewpoint is available, reducing significantly when two or more are used. In the recent paper by \cite{Kay2024}, a throughout analysis was done 
regarding the errors arising from the use of different catalogs and the subjectivity of the observers. They found that the typical difference between two independent reconstructions of the same event are $4^\circ$ in  latitude and $8^\circ$ in longitude.

The results for all CMEs studied here are presented in Table~\ref{table_gcs}. These CMEs were fast, with speeds around or higher than 1000~km~s$^{-1}$. The longitude is concentrated near the central meridian as seen from Earth, except for events 3, 4 and 13 which are closer to the limb. The majority of the events were also propagating close to the ecliptic plane, all within $30^{\circ}$ from it. The speeds shown in the table are the ones used for the cone and spheromak models. For the cone model the speed is obtained directly from the GCS fitting, for the spheromak model we used the reduced radial speeds, as derived in \cite{Scolini2019}. Regarding the CME Half-Width (HW) used in the cone model, it is calculated as:
\begin{equation}
    HW = \alpha + \arcsin{\kappa}
\label{eq_hwcone}
\end{equation}
In the case of the spheromak, what is needed is the radius at 0.1~au, which is derived as follows:
\begin{equation}
    r_{spheromak} = 21.5 \cdot \sin {\text{(HW)}}
\label{eq_hwsph}
\end{equation}

\begin{table}[ht]
\caption{3D CME parameters as derived from the GCS fitting. 1st column: event number. 2nd and 3rd columns: longitude and latitude of the CME as observed from the Earth perspective. 4th column: tilt angle with respect with the solar equator. 5th and 6th columns: aspect ratio and half-angular width of the CME, respectively. 7th column: speed of the CME. 8th column: reduced radial speed for spheromak model. All parameters are average values of measurements made at three different times and by different observers. 
}
\label{table_gcs}
\resizebox{\hsize}{!}{\begin{tabular}{c|c|c|c|c|c|c|c}
\hline
Event 	&	Longitude	& Latitude	&	Tilt	&	Ratio & Half-angular  & Speed & Speed	\\
No. & & & angle & & width &  & spheromak \\
	&	[$^\circ$]	& [$^\circ$]	&	[$^\circ$]		& & [$^\circ$] & [km~s$^{-1}$] & [km~s$^{-1}$]	\\
\hline	
1	&	W05	& S23	&	0	
& 0.40 & 30 & 1034 & 738	\\
\hline
2	&	W21	& N28	&	$-1$	
& 0.36 & 28 & 902 & 663	\\
\hline
3	&	E64	& N13	&	$-45$	
& 0.43 & 46 & 1526 & 1067	\\
\hline
4	&	E49	& N10	&	$-90$	
& 0.4 & 31 & 1351 & 965	\\
\hline
5	&	E22	& N36	&	$-84$	
& 0.38 & 40 & 2111 & 1529	\\
\hline
6	&	W33	& N20	&	$-75$	
& 0.41 & 53 & 1082 & 767	\\
\hline
7	&	E30	& S15	&	$-39$	
& 0.34 & 23 & 982 & 732	\\
\hline
8	&	W00	& S13	&	$-38$	
& 0.38 & 36 & 1160 & 840	\\
\hline
9	&	W24	& N13	&	80	
& 0.31 & 42 & 1121 & 855	\\
\hline
10	&	E14	& S07	&	$-57$	
& 0.27 & 53 & 1006 & 792	\\
\hline
11	&	E13	& S06	&	70	
& 0.21 & 28 & 947 & 782	\\
\hline
12	&	W34	& N18	&	$-77$	
& 0.38 & 57 & 888 &	643 \\
\hline
13	&	W52	& S34	&	77	
& 0.42 & 50 & 1426 & 1004	\\
\hline
14	&	W00	& N23	&	45	
& 0.55 & 38 & 1114 & 719\\
\hline
15	&	W26	& S19	&	28	
& 0.32 & 45 & 1420 & 1072	\\
\hline
16	&	W29	& S13	&	22	
& 0.37 & 50 & 1422 & 1038	\\
\hline
\end{tabular}}
\end{table}

\subsubsection{Determination of CME input magnetic parameters}
\label{subsubsection_magparam}

\begin{figure}[ht]
\centering
\resizebox{\hsize}{!}{\includegraphics{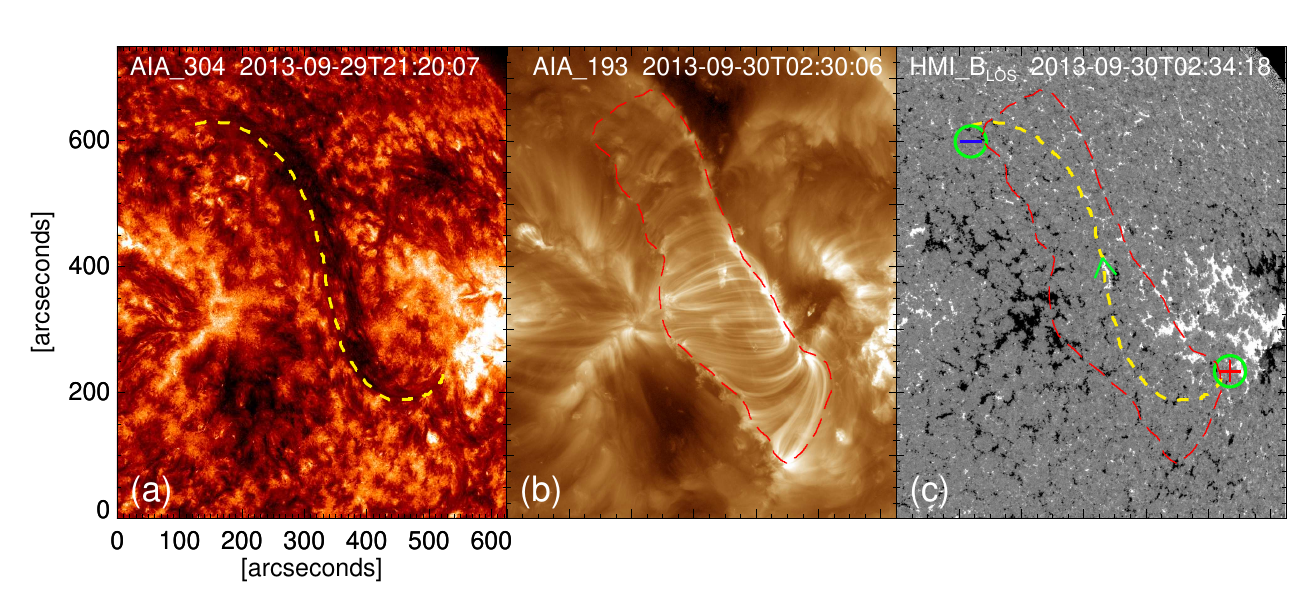}}
\caption{(a) Depicts the filament channel as indicated by the yellow dashed line in AIA \citep{Lemen2012} 304 \AA\ image. The red dashed boundary line in (b) marks the post eruption arcade (PEA) as observed in AIA 193 \AA\ image. (c) Illustrates the line-of-sight component of HMI \citep{Scherrer2012} magnetic field. The red dashed boundary and the yellow dashed line in (c) are the over-plotted PEA region and filament channel respectively. The two ends of the reverse S-shaped filament channel are marked by the green circles and the underlying magnetic polarities are shown, they indicate a northward-directed left-handed flux-rope.} \label{mag_parameters}
\end{figure}

The knowledge of magnetic parameters associated with the CMEs are crucial inputs for constraining magnetised CME models in EUHFORIA. Using the near-Sun magnetic properties of a CME as initial inputs, EUHFORIA can simulate the Sun-to-Earth evolution of a CME and provide information of its magnetic properties at 1~au. However, it is difficult to obtain a direct estimation of near-Sun magnetic properties of a CME, as the magnetic field of the solar corona cannot be reliably measured through remote-sensing observations. In this work, we use the state-of-the-art observational proxies to estimate different magnetic properties of CMEs for the 16 events studied in our Events list A, in order to constrain the model inputs of EUHFORIA.

The three magnetic parameters required to constrain a force-free magnetic flux rope (FR) are its magnetic flux, chirality and the direction of axial magnetic field \citep[e.g.,][]{Palmerio2017}. The observational techniques to constrain these parameters are detailed below. The obtained magnetic parameters for the 16 events studied are provided in Table \ref{cme_mag_parameters}.

\textbf{Poloidal flux:}
Several studies have shown that the azimuthal (i.e. poloidal) flux of magnetic FRs
formed due to reconnection is approximately equal to the low-coronal reconnection flux, which can
be obtained either from the photospheric magnetic flux underlying the area swept out by the flare ribbons
(\citealt{Longcope2007}; \citealt{Qiu2007}) or the magnetic flux underlying the Post Eruption Arcades (PEAs) (\citealt{gopalswamy2017}). The flux calculation using PEA analysis can use either line-of-sight or vector magnetograms
(\citealt{Kilpua2019a}). There is also an existing catalogue by \cite{Kazachenko2015, Kazachenko2017} that lists
flare ribbon fluxes for every flare of GOES class C1.0 and greater within $45^{\circ}$
from the central meridian,
from 2010 April until 2016 April. Magnetic field measurements are not reliable if the source
location of the flaring event lies beyond $45^{\circ}$
from the central meridian. For such cases, the empirical
relation between soft X-ray peak flux and reconnection flux can be used to estimate the poloidal flux of
the associated CME (\citealt{Tschernitz2018}; \citealt{ScoliniTemmer}).
For running the spheromak model, we need to convert the poloidal flux to toroidal flux. First, the FRED technique \citep{gopalswamy2018} is applied at 21.5~R$_\odot$ to obtain the axial magnetic field strength ($B_0$) from observations (assuming a Lunquist geometry of the magnetic cloud). The observed $B_0$ is then equated to the field strength at the magnetic axis of the spheromak model (at 21.5~R$_\odot$ - r$_{spheromak}$) \citep[][]{Sarkar2020,Sarkar_2024}. Substituting $B_0$ in equation~7 of \citealt{Verbeke2019}, the toroidal flux is obtained.

\textbf{Chirality}: One of the important properties of any FR is its helicity sign (chirality) which determines the winding direction of  the poloidal flux. In order to determine this parameter, one may apply the hemispheric helicity rule to the source active region of the CME as a first order approximation (\citealt{Pevtsov1995}; \citealt{Bothmer1998}). However, statistical studies show that the ratio of preferred helicity sign is correct only in about 60$\%$ of the cases (\citealt{Liu2014a}). Therefore, in order to confirm the chirality of the FRs, one can use further observations of pre-flare sigmoidal structures (\citealt{Rust96}), J-shaped flare ribbons (\citealt{Janvier2014}), coronal dimmings (\citealt{Webb2000}; \citealt{gopalswamy2018}), coronal cells (\citealt{Sheeley2013}) or filament orientations (\citealt{Hanaoka2017}). 

\textbf{Direction of axial magnetic field}: Orientation of the axial magnetic field of a FR can be determined by knowing the magnetic polarities of its two anchoring foot points. Analyzing the locations of the two core dimming regions or the two ends of the pre-flare sigmoidal structure, one can identify the locations of the two foot points of the FR. Thereafter, the locations of the FR foot points can be overlaid on the
line-of-sight magnetogram to determine in which magnetic polarities the FR is rooted (\citealt{Palmerio2017}).

Multi-wavelength observations of one example event (Event no. 12) are shown in Figure \ref{mag_parameters}. This is an illustration of the  observational techniques that we use to determine the magnetic parameters associated with each eruption. 

\begin{table}[ht]
\caption{Magnetic parameters of CMEs. 1st column: event number. 2nd column: helicity sign. 3rd column: axial field firection. 4th and 5th columns: Reconnection flux estimated from post-eruption arcade and from flare ribbons, ``-'' denotes inapplicable methods. 6th column: average reconnection flux.}
\resizebox{\hsize}{!}{\begin{tabular}{c|c|c|c|c|c}
\hline
Event	&	Helicity & Axial field & Reconnection &	
Reconnection  & Average  \\
	No. &	Sign	& Direction		&	flux from post   
	& flux from    & reconnection 	\\
&	&  & eruption arcade 
	& flare ribbons  & flux (poloidal) 
 \\
 &	&  & (poloidal)  
	& (poloidal)&  [$\times 10^{21}$ Mx]
 \\
 &	&  & [$\times 10^{21}$ Mx]  
	& [$\times 10^{21}$ Mx] &   
 \\
\hline	
1	&	1	& Southward	&	0.2	& - & 0.2 	\\
\hline
2	&  $-1$	& Northward	&	1.7	& 1.1 & 1.4	\\
\hline
3	&	$-1$	& Southward	&	-	& - & 9.4\tablefootmark{$\ast$}	\\
\hline
4	&	$-1$	& Southward	&	-	& - & 11.2\tablefootmark{$\ast$} 
\\
\hline
5	&	1	& Southward	&	11.1 & 10.9 & 11.0 	\\
\hline
6	&	$-1$	& North-Eastward	&	- & - & 6.8\tablefootmark{$\ast$}	\\
\hline
7	&	1	& Southward	&	1.2	& - & 1.2	\\
\hline
8	&	1	& Southward	&	14	& 5.3 & 9.7	\\
\hline
9	&	$-1$	& Northward	&	1.3	& 1.1\tablefootmark{$\star$} 
& 1.2	\\
\hline
10	&	$-1$	& Southward	&	3.8	& - & 3.8	\\
\hline
11	&	$-1$	& Northward	&	2.3	& 1.9 & 2.1	\\
\hline
12	&	$-1$	& Northward	& 5.0	& 3.2\tablefootmark{$\dagger$} & 4.1	\\
\hline
13	&	1	& Southward	&	6.5	& 5.8\tablefootmark{$\star$} & 6.1	\\
\hline
14	&	$-1$	& Northward	&	6.8	& 5.3 & 6.1	\\
\hline
15	&	$-1$	& Westward	&	0.8
& 8.7\tablefootmark{$\diamond$} & 4.8	\\
\hline
16	&	$-1$ 	& Westward	&	3.9\tablefootmark{$\diamond$} & 10\tablefootmark{$\diamond$} & 7.0	\\
\hline
\end{tabular}}
\tablefoot{
\tablefoottext{$\ast$} {Empirical relation between soft X-ray peak flux and reconnection flux is used (\cite{Tschernitz2018}; 
\cite{ScoliniTemmer}).}
\tablefoottext{$\star$} {Method from \cite{Kazachenko2017}}
\tablefoottext{$\dagger$}{Method from  \cite{Cliver2019}}
\tablefoottext{$\diamond$} {Method from \cite{ScoliniTemmer}.}
}
\label{cme_mag_parameters}
\end{table}

\section{Overview of EUHFORIA and metrics adopted}
\label{section_overview}

\subsection{Overview of EUHFORIA}

The spatial domain of EUHFORIA is divided into a coronal and heliospheric domains. The former extends from the
photosphere to 0.1~au, the latter starts there and typically extends up to 2~au. The division is done at 0.1~au because beyond that distance, the solar wind plasma is supersonic and super-Alfvénic which means that no information is traveling towards the Sun \citep{Pomoell2018}. Both models are in principle independent of each other, and different models could be used as far as the correct coupling is assured. An adaptation of the WSA model \citep[WSA,][]{2003Arge} is used for the coronal part.  The heliospheric model of EUHFORIA is a three-dimensional time-dependent magnetohydrodynamics simulation that solves the ideal MHD equations and self-consistently models the propagation, evolution and interaction of solar wind and CMEs.

For producing a solar wind simulation, only a synoptic magnetogram is needed as input. When CMEs are inserted into the model, the CME input parameters are then also needed. They will depend on the CME model used. As output of its heliospheric model, EUHFORIA provides plasma and magnetic field quantities at any location of its domain. 

Before launching the CME into EUHFORIA's heliospheric domain, the background solar wind  can be optimised in order to reproduce the correct ambient solar wind in which the CME will be injected. For this purpose, EUHFORIA is run with different Global Oscillation Network Group (GONG) magnetograms\footnote{http://gong.nso.edu/data/magmap/QR/} close to the CME launch time. The one that best reproduces the observed solar wind at 1 AU before the ICME arrival, is used. To pick the best magnetogram for modeling the solar wind before an ICME arrival, we carefully searched through available options. We started with the closest magnetogram to the CME launch time and ran simulations. If the initial magnetogram failed to yield satisfactory results, we systematically reviewed others at 6-hour intervals until finding the most suitable one. Most of the magnetograms were selected within three days of the CMEs (with a few up to six days). 
This method ensures that the chosen magnetogram drives EUHFORIA to accurately represent the solar wind conditions, enabling realistic ICME propagation simulations.
This is an optional optimization, here it was done for our Event list A, the selected magnetogram for each of the runs can be found in Table~\ref{table_eventslist}.  
The CMEs are then inserted into EUHFORIA at 0.1~au  using the cone and spheromak models. Additionally, CMEs occurred in the days preceding Events 5, 6, and 14 were included in the simulation in order to create more realistic conditions for the main CMEs. Further details on these events can be found in Appendix \ref{supp_ev}. For Event list B, we used always the magnetogram taken six hours before the CME, this would be the situation closest to a real time CME forecasting scenario.

\subsubsection{The cone CME model}
\begin{figure}[ht]
\centering
\resizebox{\hsize}{!}{\includegraphics{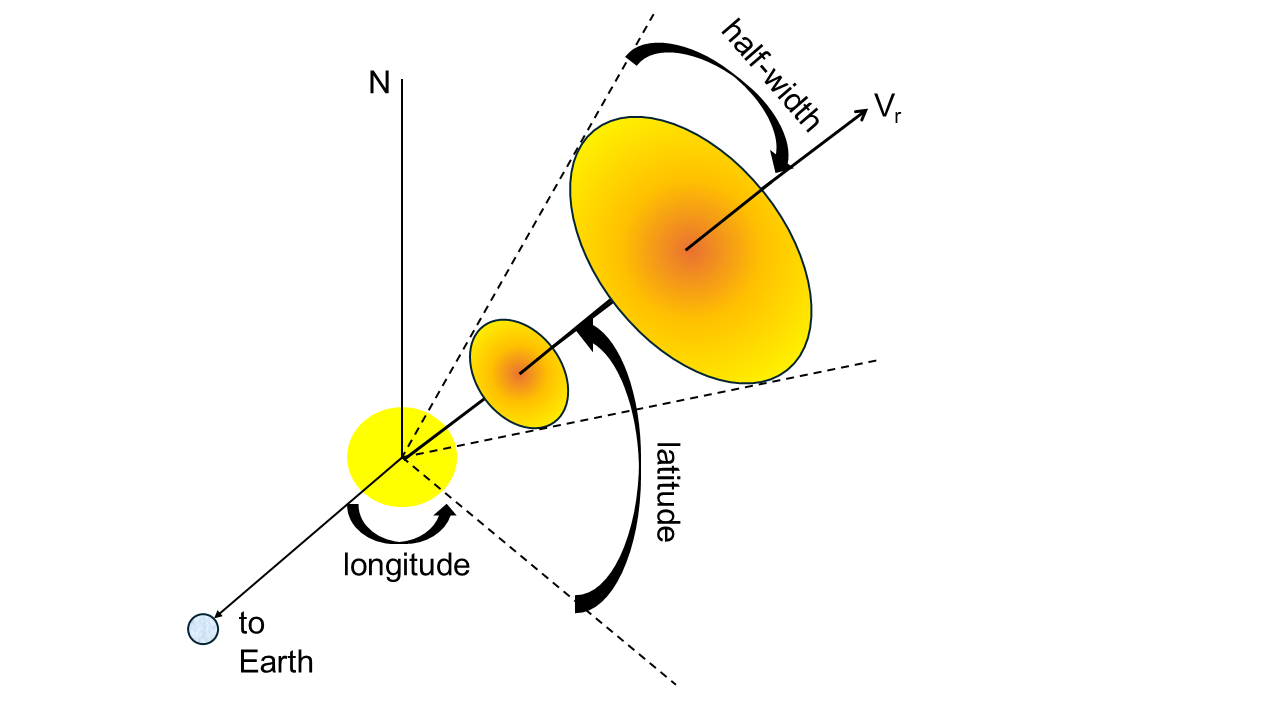}}
\caption{Schematics of the cone model. Adapted from \cite{Dewey2015} } 
\label{fig_cone}
\end{figure}

The cone CME model treats the ejecta as a hydrodynamic (velocity and density) pulse injected at the inner radial boundary of the simulation domain. It is characterized 
by a self-similar expanding geometry \citep{Xie2004,Odstrcil1999}, and it does not contain a prescribed internal magnetic field. This does not mean that the internal magnetic field of the ICME will be zero, the pulse is injected into the solar wind and will interact with its magnetic field. Due to its simplicity, the cone CME model has been widely used in 3D MHD simulations in the past decades. 
Figure \ref{fig_cone} shows a graphic representation of the cone model, together with the parameters normally required as input for EUHFORIA (derived from coronagraphic observations, Section \ref{subsubsection_geoparam}). It represents a CME that propagates radially outward for the Sun and expands while keeping constant angular width.

\subsubsection{The spheromak model}
\begin{figure}[ht]
\centering
\resizebox{\hsize/3*2}{!}{\includegraphics{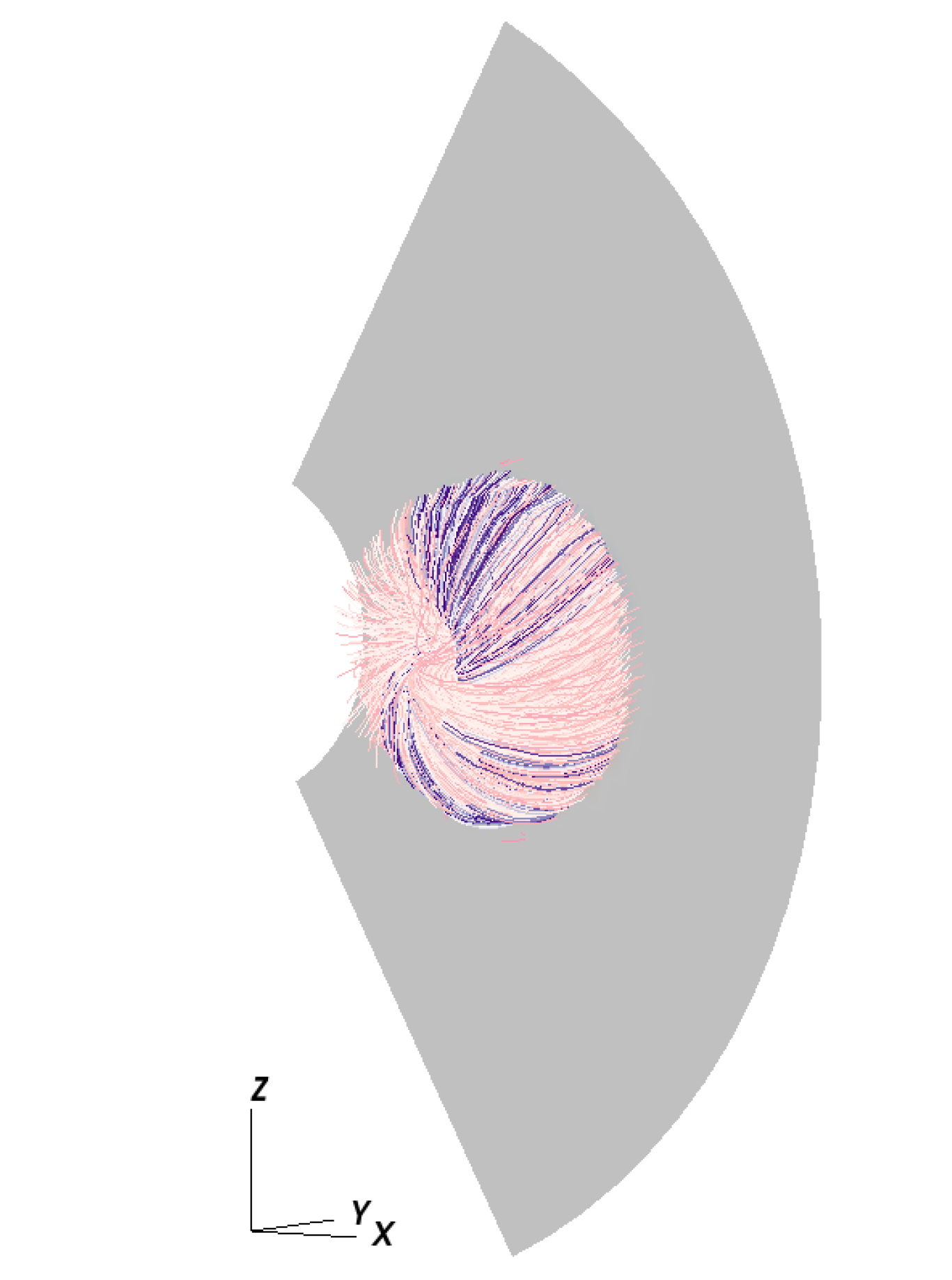}}
\caption{Magnetic field lines depicting the structure of the spheromak CME model. The grey plane shows the meridional HEEQ y=0-plane. From \cite{Verbeke2019}.} 
\label{fig_spheromak}
\end{figure}

The linear force-free spheromak CME model was implemented in EUHFORIA \citep{Verbeke2019} in order to allow the possibility of inserting CMEs into the heliospheric domain that contain a structured internal magnetic field, with a toroidal-like flux rope structure in this case. 
Figure~\ref{fig_spheromak} presents a visualization of the magnetic field structure of the model, from \cite{Verbeke2019}. In this case, on top of the geometric parameters, a set of magnetic parameters will be needed (Section \ref{subsubsection_magparam}). The CME in this model is considered to be a sphere upon the time of its injection, it is launched outward following the latitude and longitude directions specified as input. The magnetic field structure is defined in a local spherical coordinate system, with origin at the centre of the spheromak and symmetry in the azimuthal direction. 

\subsection{Metrics description} 
\label{Section_Metrics} 
This section describes the list of metrics used for the comparison of EUHFORIA output with the in situ data at 1~au. 

Within the field of space weather forecasting, it is crucial to assess the reliability and accuracy of predictive models. A widely employed tool for this purpose is the contingency table, see e.g. \cite{Verbeke2019arrival}. Events are categorized into four classes: hits (indicating correct prediction of space weather event), misses (representing instances where the model fails to predict a space weather event), false alarms (signifying predictions of the space weather event that does not occur), and correct rejections (denoting correct prediction of quiet period). This analysis can be also conducted using different time intervals,  which are needed to specify the maximum time allowed for the hit to occur (e.g. 24 hours), between observed and predicted space weather events.


%

\begin{table}[h!]
\caption{Skill Scores, based on \cite{Verbeke2019arrival} }
\centering
\resizebox{\hsize}{!}{\begin{tabular}{c|c|c} 
\hline
  Skill Score & Equation & Perfect Score  \\ & &  ~\\[-1.8ex] 
 \hline
 ~\\[-1.8ex]
 Probability of Detection (POD)  & $\frac{H}{H + M}$ & 1 \\ & & ~\\[-2ex] \hline
 ~\\[-1.8ex]
 Success Ratio (SR)  & $\frac{H}{H + FA}$ & 1 \\ & &  ~\\[-1.8ex]
\hline ~\\[-1.8ex]
Bias Score (BS)  & $\frac{H + FA}{H + M}$ & 1 \\ & &  ~\\[-1.8ex]
\hline ~\\[-1.8ex]
Critical Success Index (CSI)   & $\frac{H}{H + M + FA}$ & 1 \\ & &  ~\\[-1.8ex]
\hline ~\\[-1.8ex]
Accuracy (Ac)  & $\frac{H + CR}{Total}$ & 1 \\ & &  ~\\[-1.8ex]
\hline ~\\[-1.8ex]
False Alarm Rate (POFD)  & $\frac{FA}{H + FA}$ & 0 \\ & &  ~\\[-1.8ex]
\hline ~\\[-1.8ex]
Hanssen and Kuipers discriminant  & $HK = POD - POFD$ & 1 \\[0.8ex]
\hline
\end{tabular}}
\label{table_metrics2}
\end{table}
Based on the contingency table, some classic metrics can be calculated, such as the probability of detection (POD), the success ratio (SR), the bias score (BS), the critical success index (CSI), the accuracy (Ac), the false alarm rate (POFD) and the Hanssen and Kuipers discriminant. These metrics are listed in Table \ref{table_metrics2} and are complementary to each other. These metrics can be visualized in a single figure for clarity, known as the performance diagram (e.g. see \cite{Verbeke2019arrival}). 

In addition to the previously mentioned metrics, we also employ a set of metrics specifically calculated within the ICME interval, focusing on the cases where ICMEs arrival is predicted by EUHFORIA and observed (Hit events). The metrics applied in this case were the mean Error (ME), the mean absolute error (MAE), the mean square error (MSE), the standard deviation (SD) and the root-mean-square error (RMSE). These metrics are computed based on key parameters such as the CME arrival time and $K_p$ index. Further details regarding the computation and interpretation of these metrics can be found in \cite{Verbeke2019arrival}.

\section{Results - Event list A: Selected Events}
\label{section_resultsA}

\subsection{Comparison of arrival times between simulated and real CMEs} 
\label{subsection_arrival}
The arrival times for each simulation output of the Events list A are shown in Table~\ref{table_eventslist}. The arrival time estimates are compared with the observed arrival times, while the estimates of the ICME properties from the simulations are compared with measured in situ data obtained from the OMNI database. In particular, we have focused on comparison of the solar wind bulk speed and magnetic field. The geomagnetic impact is estimated using the $K_p$ index. In Appendix \ref{events_plots}, we provide plots of each event, including observations and results from simulations.

\begin{figure}[ht]
\centering
\resizebox{\hsize}{!}{\includegraphics{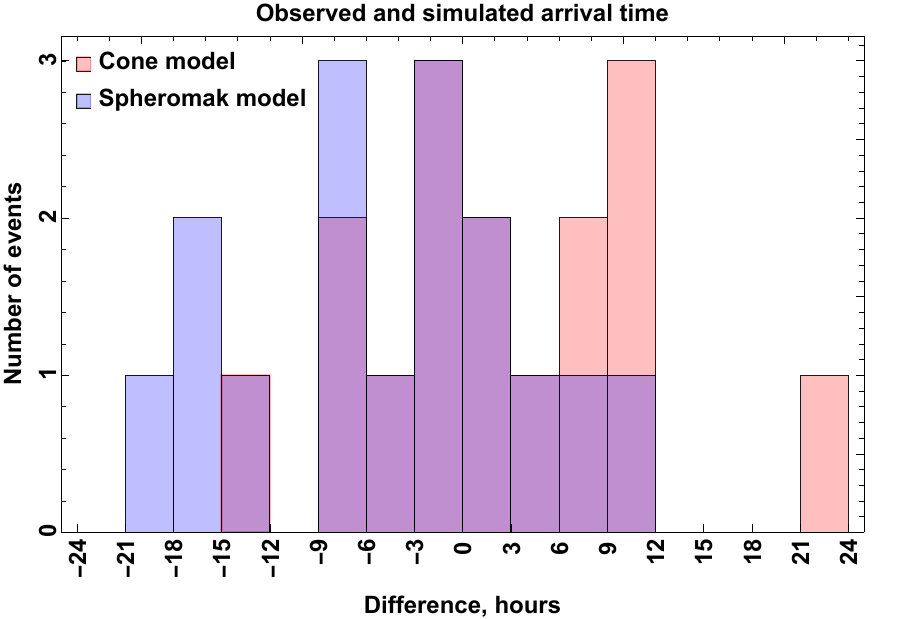}}
\caption{Histogram of difference between observed and simulated arrival time for the cone (pink) and spheromak (blue) models. Overlapping instances are represented in purple.} \label{arrival_time_cone_spheromak}
\end{figure}

Figure~\ref{arrival_time_cone_spheromak} displays a histogram of time difference between observed and simulated ICME arrival time for cone and spheromak runs. Cone CMEs tend to arrive earlier than spheromak ones.
EUHFORIA correctly predicts that all 16 events arrived to the Earth. 
All the ICMEs arrive within 24 hours, with RMSE values of around 9 hours, for both cone and spheromak. This is a good result, close to the values of 10 or 12 hours for Enlil \citep{Odstrcil2003} observed in previous studies \citep{Mays2015, Riley2018, Riley2021, Wold2018}. If we use $\pm$~12h as a limit for evaluating the probability of detection of the model, we obtain that the the cone model (14 hits out of 16 events) scores slightly higher than the spheromak model (12 hits out of 16 events).
In Table \ref{table:ConeSpherArrival} we show the RMSE and SD for both cone and spheromak arrival times, for the CMEs that arrive within 12 hours time frame and within 24 hours (all ICMEs), with respect to the real observed times. All values in this table are below 10 hours. 100\% of the ICMEs arrived within 24 hours, and 93\% within 18 hours.
\begin{table}[ht]
\caption{Cone and spheromak models Root Mean Square Error (RMSE) and Standard deviation (SD) in arrival time, measured in hours.}
\begin{centering}
\begin{tabular}{c|c|c|c|c}
\hline
 & RMSE & SD  & RMSE  & SD \\
 & 12 h & 12 h & 24 h & 24 h \\ 
 \hline
 Cone model   & 6.75  & 6.49 & 9.35  & 9.08  \\
 \hline
 Spheromak model & 5.49 & 5.43 & 9.53  & 8.29  \\
\hline
\end{tabular}
\end{centering}
\label{table:ConeSpherArrival}
\end{table}

\subsection{Comparison of speed and geomagnetic impact, between simulated and real CMEs} 
\label{subsection_VandGeo}

\begin{figure}[ht]
\centering
\resizebox{0.7\hsize}{!}{\includegraphics{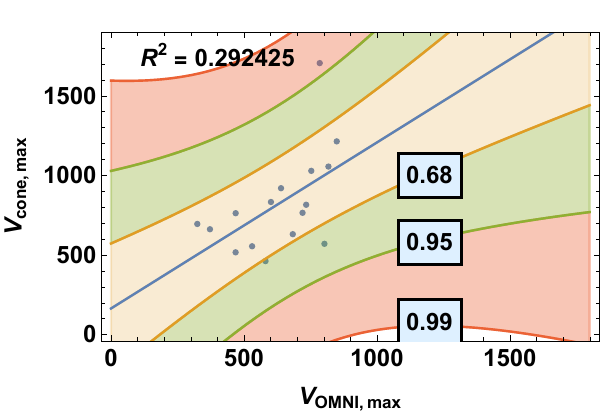}}
\resizebox{0.7\hsize}{!}{\includegraphics{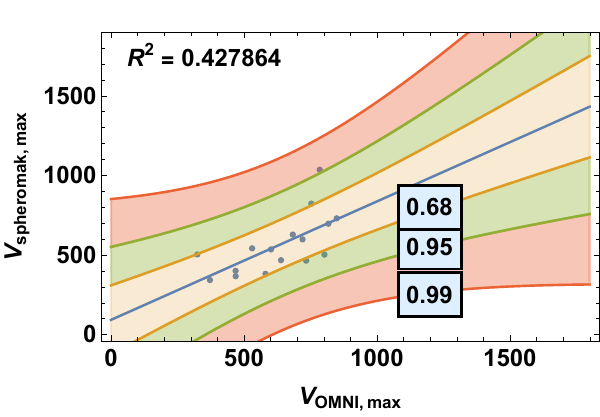}}
\caption{Correlation between the maximum value of OMNI speed and the maximum value of cone (top) and spheromak (bottom) speed. The $68\%$, $95\%$, and $99\%$ confidence bands are marked in yellow, green, and red, respectively.} \label{correlation_cone_3}
\end{figure}

We evaluated correlation factors for the maximum value of speed between observed and simulated ICMEs in Events list A. Results are shown in Figure~\ref{correlation_cone_3} for the cone and spheromak runs. Both cone and spheromak runs have a positive linear correlation between observed and simulated maximum value of ICME speed as expected. The spheromak model has a better performance when predicting the maximum speed. It is noteworthy, however, that there remains a significant spread in values; for example, an in-situ speed of $\sim 800$ km~s$^{-1}$ may correspond to predictions ranging from 400 to 1000 km~s$^{-1}$.


The impact of a CME on Earth can be estimated using geomagnetic indices. Here we use  the  $K_p$ index that measures global geomagnetic activity. In its current version, EUHFORIA uses the simulation output in the solar wind upstream of the Earth's magnetosphere to calculate the $K_p$-index, based on Eq. \ref{Kpeq}, by \cite{2008Newell}.

\begin{equation}
    K_p = 0.05 + 2.244 \cdot 10^{-4} \frac{d \Phi_{MP}}{d t} + 2.844 \cdot 10^{-6} n^{1/2} V^2, 
    \label{Kpeq}
\end{equation}

Here $d \Phi_{MP} / dt$ is the rate at which magnetic flux is opened at the magnetopause, defined by
\begin{equation}
\nonumber
\frac{d \Phi_{MP}}{d t} = V^{4/3} B^{2/3} \sin^{8/3} (\theta_c/2),
\end{equation}
$V$ is solar wind speed, $n$ is density, $B$ is the magnetic field magnitude and $\theta_ c = \arctan (B_y / B_z)$. 
We compared $K_p$ indices calculated for cone ($K_{p,\text{cone}}$) and spheromak ($K_{p,\text{spheromak}}$) models with the corresponding observed $K_p$ values ($K_p,\text{observed}$). We also compute the $K_p$ values using Eq.~(\ref{Kpeq}) based on observed solar wind parameters from the OMNI database ($K_{p,\text{calculated}}$). In this way, apart from comparing $K_{p,\text{cone}}$ and $K_{p,\text{spheromak}}$ with $K_{p,\text{observed}}$, we can also compare $K_{p,\text{cone}}$ and $K_{p,\text{spheromak}}$ with $K_{p,\text{calculated}}$, which gives us a better way of comparing the impact of the different CME models independent of the empirical $K_p$ formula used.

In order to evaluate how well EUHFORIA predicts the geoeffectiveness of ICMEs, we  classify our events in three groups, depending on their $K_p$-index 
\begin{itemize}
    \item No storm observed, i.e $K_p \leq 4$
    \item Moderate to strong storm $5 \leq K_p \leq 7$
    \item Severe or extreme storm $K_p \geq 8$
\end{itemize}

The correctness in the prediction of the $K_p$-index by EUHFORIA is assessed using the following criteria: a prediction from the cone and spheromak models is considered a Hit if the maximum predicted $K_p$ value falls within the same interval as $K_{p,\text{observed}}$ (or $K_{p,\text{calculated}}$), a False Alarm if it overestimates the geoeffectiveness of the storm (higher $K_p$ than observed), and a Miss if it underestimates it (lower $K_p$ than observed).  There is no Correct Rejection in this case. The maximum value of $K_p$ for each of the selected events can be found in Table \ref{table:Kpmaxvalue}. 
Furthermore, Table \ref{table:MetricsKp} provides metrics based on the probability of detection (POD), the success ratio (SR), the standard deviation (SD) and the root mean square error (RMSE).


\begin{table}[h!]
\caption{Maximum values of $K_p$:  $K_{p,\text{observed}}$ (Column 2), EUHFORIA $K_{p,\text{cone}}$ (Column 3), EUHFORIA $K_{p,\text{spheromak}}$ (Column 4), and $K_{p,\text{calculated}}$ (Column 5) using Eq. \ref{Kpeq} and the observed solar wind parameters from OMNI.}
\begin{centering}
\begin{tabular}{c|c|c|c|c}
 \hline
 Event No. & $K_p$ & $K_p$  & $K_p$ & $K_p$       \\
 & observed & cone & spheromak &  calculated \\
 \hline
 1  & 7.67  & 8.87 &  7.34 &  8.14 \\ \hline
 2  & 5.67 & 3.84 &  3.09 &   6.38 \\ \hline
 3 & 2.0  & 6.98  & 2.51 &  3.26\\ \hline
 4 & 6.33  & 5.10 & 3.72 &  8.12 \\ \hline
 5  & 8.0 & 9.0 &  5.83 &  8.54\\ \hline
 6  & 6.33  & 9.0  & 3.62  & 7.88 \\ \hline
 7  & 2.33  & 3.81 & 2.29  & 2.51 \\ \hline
 8  & 7.0 & 7.72 &  9.0 &  8.55\\ \hline
 9  & 6.67 & 6.41 &  3.45  & 5.17 \\ \hline
 10  & 6.67 & 9.0 &  9.0 & 8.78\\ \hline
 11  & 3.33 & 3.51 & 3.93  & 4.17   \\ \hline
 12 & 7.67  & 9.0 & 2.44  & 8.55\\ \hline
 13 & 2.67  & 8.99 & 2.34   & 2.22\\ \hline
 14 & 6.33  & 9.0 & 8.15 &   8.23 \\ \hline
 15 & 3.67  & 9.0 & 5.65 & 5.87 \\ \hline
 16 & 8.33  & 9.0 &  6.29 & 8.63 \\ \hline
\end{tabular}
\end{centering}
\label{table:Kpmaxvalue}
\end{table}

If we first compare the $K_{p,\text{observed}}$ and the $K_{p,\text{calculated}}$ using the measured solar wind parameters in Eq. \ref{Kpeq} (i.e. without using the predictions from the EUHFORIA CME models), we can see that they fall on the same $K_p$ interval for 10 events. For the remaining 6 cases, the $K_{p,\text{calculated}}$ overestimates the CME impact on the Earth (events 4, 6, 8, 10, 14, 15, Success Ratio = 0.63). The $K_{p,\text{calculated}}$ has an RMSE of 1.31 when compared to the observed one, with SD = 0.96 (Table \ref{table:MetricsKp}). 

The cone model predicted correctly the observed geomagnetic impact of the CMEs for 8 events (POD = 0.88), it overestimated the impact for 7 events (3, 6, 8, 10, 13, 14, 15) and underestimated it for event 2. The RMSE amounts to 2.8, with a SD = 2.20. If we now compare the $K_{p,\text{cone}}$ with the $K_{p,\text{calculated}}$ (instead of the $K_{p,\text{observed}}$), then the correct predictions increase to 11 events, with an underestimation for 2 events (2, 4) and overestimation for 3 events (3, 13, 15). The RMSE error is then improved to 2.40, with SD = 2.25.

\begin{table}[h!]
\caption{ Metrics used for estimating the performance of $K_{p,\text{cone}}$ and $K_{p,\text{spheromak}}$ with respect to $K_{p,\text{observed}}$ and $K_{p,\text{calculated}}$. We also present the comparison between $K_{p,\text{calculated}}$ and $K_{p,\text{observed}}$.}
\begin{centering}
\begin{tabular}{c|c|c|c|c}
 \hline
 & POD & SR  & SD  & RMSE  \\
 \hline
 $K_{p,\text{calculated}}$ vs.$K_{p,\text{observed}}$& 1.0 & 0.63 & 0.96 & 1.31 \\ \hline
 $K_{p,\text{cone}}$ vs. $K_{p,\text{observed}}$ & 0.88  & 0.53 & 2.2  & 2.8  \\ \hline
 $K_{p,\text{spheromak}}$ vs. $K_{p,\text{observed}}$ & 0.42 & 0.56 & 2.18  & 2.30  \\
\hline
 $K_{p,\text{cone}}$ vs. $K_{p,\text{calculated}}$ & 0.84  & 0.78 & 2.26  & 2.40  \\ \hline
 $K_{p,\text{spheromak}}$ vs. $K_{p,\text{calculated}}$ & 0.5 & 1.0 & 1.91  & 2.54 \\  \hline
\end{tabular}
\end{centering}
\label{table:MetricsKp}
\end{table}
 
The spheromak model performs worse in predicting the level of $K_p$ activity than the cone model. It predicted $K_p$ index within the correct interval (compared to $K_{p,\text{observed}}$) for 5 events, underestimated it for 7 events (2, 4, 5, 6, 9, 12, 16) and overestimated it for 4 events (8, 10, 14, 15), with an SD~=~2.18 and RMSE~=~2.30.  However, comparison with the $K_{p,\text{calculated}}$ shows better results, it estimated the correct geomagnetic storm level for 8 events, and underestimated for 8 events (1, 2, 4, 5, 6, 9, 12, 16) with an SD~=~1.93, RMSE~=~2.54. These numbers show that the cases where the $K_{p,\text{spheromak}}$ was overestimated, were mainly caused by overestimation in the $K_p$ formula (all overestimated cases disappear when we compare with the calculated $K_p$). 

\begin{figure}[ht]
\centering
\resizebox{\hsize}{!}{\includegraphics{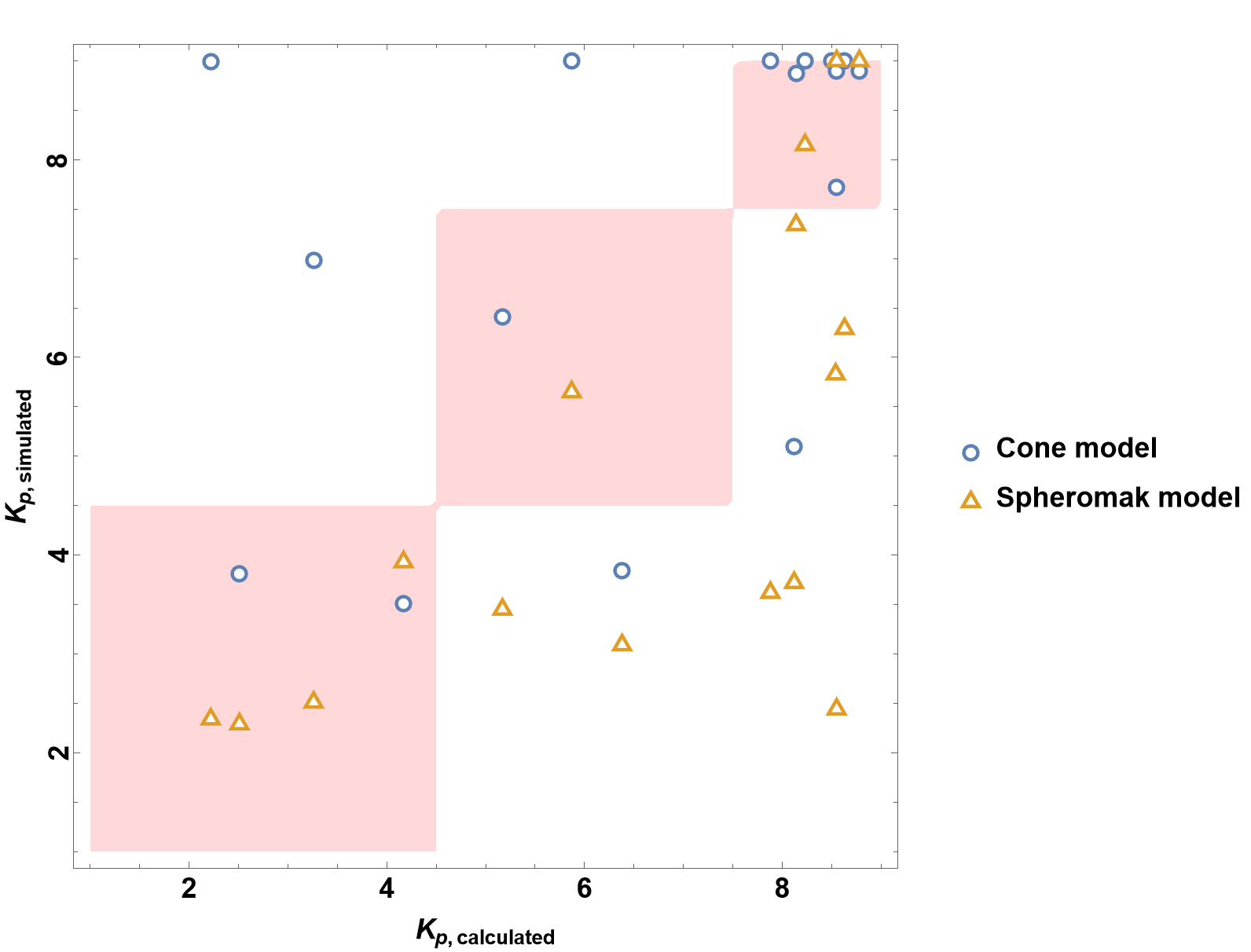}}
\caption{Comparison of simulated $K_p$-index value with $K_{p,\text{calculated}}$ for cone and spheromak models. } \label{kp-index-comparison}
\end{figure}

The results of the comparison of simulated vs. $K_{p,\text{calculated}}$ for cone and spheromak models are shown in Figure \ref{kp-index-comparison}.


For 3 events (5, 9, 12), the spheromak model underestimates the $K_p$ index while the cone model predicts the value in the correct range. The opposite occurs for 2 events (3, 13), the cone model overestimates the value, while the spheromak model captures it correctly. 
This can be explained by the simulated solar wind speed and density in the cone model having higher values than for the spheromak model for these cases, overcompensating the lower magnetic fields encountered by the cone model. For events (3,13), the cone model overestimates the speed and density by a great extent, hence, abnormally overestimating the $K_p$ index. 
One more possible reason for the cases when the cone performs better than the spheromak is that for the latter, even when getting the speed at arrival correctly, an incorrect rotation of the $B_z$ field could influence the resulting $K_p$.

\section{Results - Event list B: Evaluation of CME arrival forecast}
\label{section_resultsB}

In order to validate the capabilities of EUHFORIA in forecasting the correct arrival at the Earth, we employ Events list B and the contingency tables approach, as described in Section ~\ref{Section_Metrics}. In this case, Hit corresponds to events for which a CME arrival is both predicted and detected by instruments. False Alarm occurs when a CME arrival is predicted, but the ICME is not observed to arrive. Miss represents cases where a CME arrives despite lacking a prior prediction. Finally, Correct Rejection is the case of an event, for which the CME is neither predicted to arrive nor observed.

\begin{figure}[ht]
\centering
\resizebox{\hsize}{!}{\includegraphics{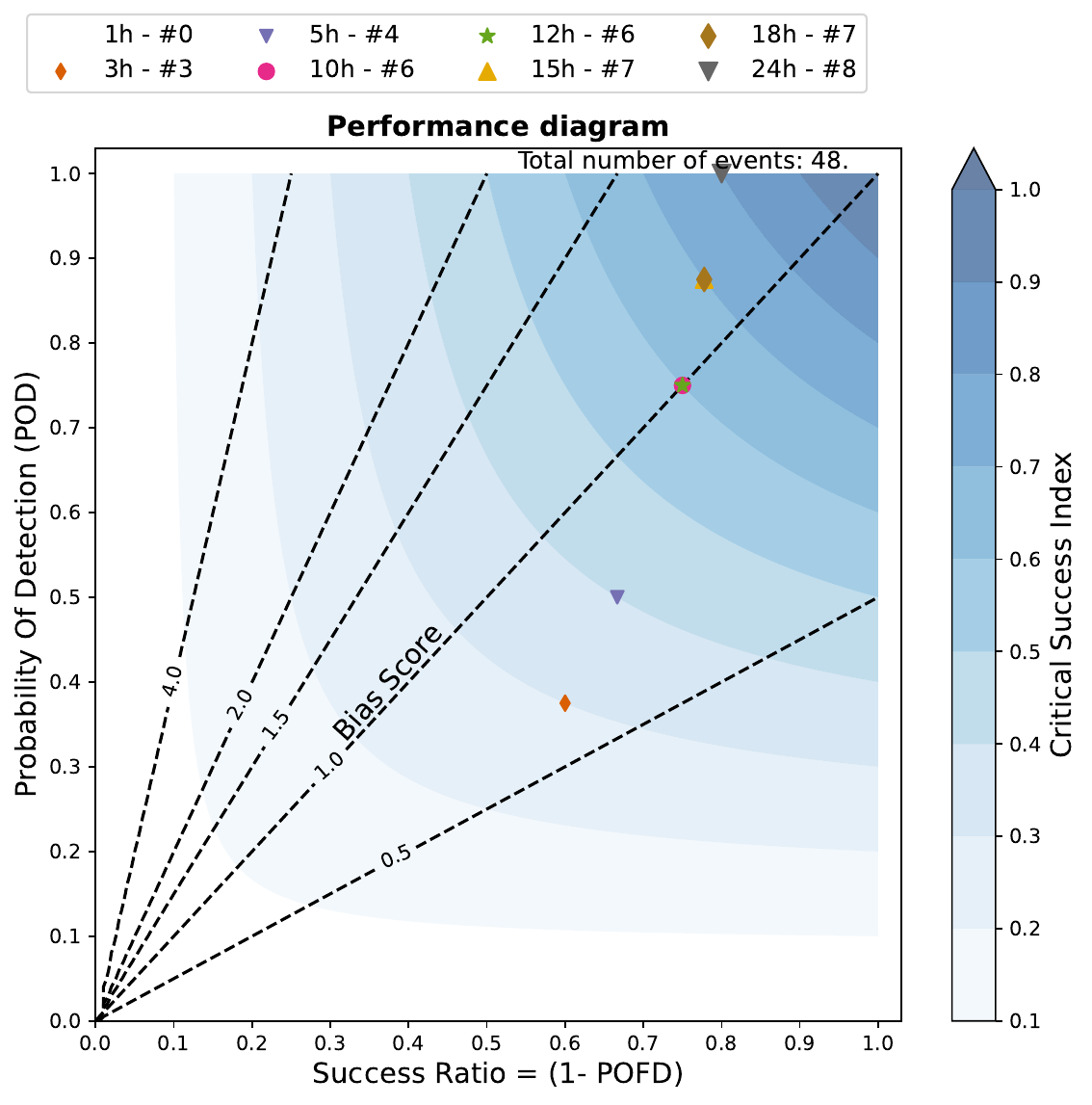}}
\caption{Performance diagram for the events in Section \ref{subsubsection_eventsB}.
The dashed diagonal lines correspond to lines of equal Bias Score, while the blue contours correspond to equal Critical Success  Index.  The coloured dots are used to mark different time intervals allowed for the ICME to arrive. The shortest interval is 1 hour (no ICME arrivals) and the largest one is 24 hours (for which all the ICMEs arrive).}
\label{performance_2010_2012}
\end{figure}

A complete summary including all the classic metrics is given in Table \ref{table:ConeSpherArrival2010-2012}. Figure~\ref{performance_2010_2012} shows some of the metrics used in this study in a graphical way. EUHFORIA performs well, with a 75\% Probability of Detection (POD) in a 12~hours time window (with 4~hours RMSE and SD) and a 100\% POD in a 24~hours time window (with 9~hours RMSE and 7.6~hours SD). The Bias Score (BS) is balanced at unity for 12-hours and showing a slight prevalence of false alarms over misses in the 24~hours time window. The Critical Success Index (CSI) improves when moving from 12 to 24~hours time window, as the number of correctly forecasted events increases. The Accuracy (Ac) is very high in both time windows, as this metric depends mostly on the correct rejections, in which EUHFORIA provides a very good performance. The Probability of False Alarm rate (POFD) remains very low for both time windows. The Hanssen and Kuipers discriminant is very close to the POD (since POFD is very small for both time windows). In summary, the classic metrics show that EUHFORIA is very good at correctly predicting the arrival of ICMEs.

\begin{table}[h!]
\caption{Metrics for the set of CMEs in 2010 and 2012. The table shows the Root Mean Square Error (RMSE) in arrival time, the Standard deviation (SD) in arrival time, the Hit rate or Probability of Detection (POD), the Success Ratio (SR), the Bias Score (BS), the Critical Success Index (CSI), the Accuracy (Ac), the False Alarm Rate (FAR) or Probability of False Detection (POFD), and the HK discriminant (HK).}
\begin{centering}
\resizebox{\hsize}{!}{
\begin{tabular}{c|c|c|c|c|c|c|c|c|c}
\hline
   Interval & RMSE & SD & POD & SR & BS & CSI & Ac & POFD & HK \\
 \hline
   12 h  & 4.09  & 3.84  & 0.75 & 0.75 & 1.0 & 0.6 & 0.91 & 0.05 & 0.7 \\ \hline
   24 h  & 9.21  & 7.62  & 1.0 & 0.8 & 1.25 & 0.8 & 0.95 & 0.05 & 0.95 \\
 \hline
\end{tabular}}
\end{centering}
\label{table:ConeSpherArrival2010-2012}
\end{table}

\section{Conclusions}
\label{Summary}
In this paper, we presented a detailed analysis of the validation of the cone and spheromak CME models in the framework of EUHFORIA. We used two datasets to validate the geoeffectiveness and arrival time predictions respectively. The first dataset was composed of 16 CMEs with an ICME counterpart that arrives to the Earth. In this way, we can compare the in situ solar wind and magnetic field parameters as measured at the L1 point. Furthermore, we compared their geomagnetic impact by means of the $K_p$-index. Input parameters for all the CMEs were constrained from observations and used to run EUHFORIA. 

Results show that cone CMEs tend to arrive earlier than spheromak ones. All the ICMEs arrive within 24 hours. The RMSE value is 9 hours, for both cone and spheromak. These values are comparable to the 10 hours found by \cite{Mays2015,Riley2021, Wold2018} for similar MHD models. If we focus on the 12h time frame, the probability of detection of the CME for cone numerical simulations is slightly higher than for spheromak runs, with 14 events out of 16 arriving, compared to 12 events for spheromak runs. Regarding the obtained ICME speeds, both cone and spheromak runs have a positive correlation between observed and simulated maximum value of ICME speed as expected, with a large spread. The spheromak model does a slightly better job at predicting the maximum speed. 

We estimated the impact of an ICME at Earth by means of the $K_p$-index. We compared $K_p$ indices calculated for cone ($K_{p,\text{cone}}$) and spheromak ($K_{p,\text{spheromak}}$) models both with the  observed $K_p$ values ($K_{p,\text{observed}}$) and with the calculated $K_p$ values using Eq. (\ref{Kpeq}) based on observed solar wind parameters from the OMNI database ($K_{p,\text{calculated}}$). The comparison between the $K_{p,\text{calculated}}$ and $K_{p,\text{observed}}$ has an RMSE of 1.31. 
The cone model correctly forecasted the $K_{p,\text{calculated}}$ range for 11 out of 16 events (69\%) with an RMSE of 2.4. If we instead use the $K_{p,\text{observed}}$, the correct forecast reduces to 8 out of 16 events (50\%) and the RMSE value reaches 2.8. The spheromak model forecasted $K_p$ within the correct interval for 5 events only (31\%) when compared with $K_{p,\text{observed}}$, with an RMSE value of 2.3. However, comparison with $K_{p,\text{calculated}}$ increased the correct forecast to 8 events (50\%) with an RMSE of 2.54. The cases where the $K_{p,\text{spheromak}}$ were overestimated, were caused by an overestimation introduced by the $K_p$ formula. In principle, one would have expected the spheromak model to provide a better estimation of $K_p$, but this was not the case here. One possible reason could be edge encounters. Because of its compact spherical
shape and lack of CME legs, the spheromak has difficulties to model the events when ICMEs impact Earth with their flanks. One more possible reason for the cases when the cone performs better than the spheromak is that for the latter, even when getting the speed at arrival correctly, an incorrect rotation of the $B_z$ field could influence in the $K_p$ result. In the case of the cone model, there is no prescribed internal magnetic field within the CMEs. Nevertheless, a magnetic field is present there, as everywhere else in the EUHFORIA simulation domain. This magnetic field is affected by the CME plasma dynamics (e.g. compression), creating a distinctive magnetic field inside the ICME, compared to the background solar wind. New CME flux rope models are currently being tested with EUHFORIA, an improvement in this important aspect is expected in the future. Recent developments in flux rope CME models include the implementation of the  Flux Rope in 3D (FRI3D) model \citep{Maharana2022}  and a toroidal CME model \citep{Linan2024} into EUHFORIA.

In order to evaluate EUHFORIA forecasts in terms of arrival (hits or misses) of CMEs, we used a second (larger) dataset of CMEs for which their arrival at the Earth was not a condition. We took all frontsided CMEs wider than 60$^{\circ}$ and faster than 350 km~s$^{-1}$, during eight months. Four months were taken during solar minimum (June - September 2010) and the other four during solar maximum (June - September 2012). The final dataset contained 48 CMEs that were simulated with EUHFORIA cone model in order to predict Earth arrival or misses.The spheromak model was not used in this study, as we were mainly interested in evaluating the arrival of ICMEs and not their internal magnetic field configuration. EUHFORIA showed a 75\% probability of detection in a 12 hour time window (with 4 hours RMSE) and 100\% probability of detection in a 24 hours time window (with 9 hours RMSE). In this dataset, the events were not carefully selected as the previous one of 16 CME-ICME pairs, and thus shows that EUHFORIA can perform well even when CMEs are not handpicked.

These results validate the use of cone and spheromak CMEs in a real-time space weather forecasting manner, even with its simpler cone CME model. EUHFORIA is currently being used for such purposes by the space weather forecast team at the Royal Observatory of Belgium.

\begin{acknowledgements}
EUHFORIA was created as a joint effort between KU Leuven and the University of Helsinki and was developed further by the the project EUHFORIA 2.0, a European Union's Horizon 2020 research and innovation programs under grant agreement No 870405. These results were also obtained in the framework of the projects C14/19/089 (C1 project Internal Funds KU Leuven), G.0D07.19N and G.0B58.23N (FWO-Vlaanderen), 4000134474 SIDC Data Exploitation (ESA Prodex-12), and BELSPO projects BR/165/A2/CCSOM and B2/191/P1/SWiM. The Royal Observatory of Belgium team thanks the Belgian Federal Science Policy Office (BELSPO) for the provision of financial support in the framework of the PRODEX Programme of the European Space Agency (ESA) under contract numbers 4000134088, 4000112292, 4000134474, and 4000136424. The OMNI data were obtained from the GSFC/SPDF OMNIWeb interface at \url{https://omniweb.gsfc.nasa.gov}. SOHO LASCO CME catalog is generated and maintained at the CDAW Data Center by NASA and The Catholic University of America in cooperation with the Naval Research Laboratory. SOHO is a project of international cooperation between ESA and NASA.
RS acknowledges support from the Academy of Finland Grant 350015. 
EA acknowledges support from the Academy of Finland/Research Council of Finland (Postdoctoral Researcher grant number 322455 and Academy Research Fellow grant number 355659). 
JP acknowledges support from the Academy of Finland project SWATCH (343581).
ES research was supported by an appointment to the NASA Postdoctoral Program at the NASA Goddard Space Flight Center, administered by Oak Ridge Associated Universities under contract with NASA. 
We acknowledge the Community Coordinated Modeling Center (CCMC) at Goddard Space Flight Center for the use of StereoCAT, https://ccmc.gsfc.nasa.gov/analysis/stereo/.
\end{acknowledgements}

%
%

\bibliographystyle{aa}
\bibliography{biblio}  

\begin{appendix}
\section{Supplementary events for improving background Solar Wind modeling.}
\label{supp_ev}
Table \ref{extra_cme} provides detailed information on additional events that occurred in the the days preceding Runs 5, 6, and 14. Those events were included in the simulation in order to create more realistic conditions for the main CMEs. It is important to note, that the CMEs listed in the supplementary table \ref{extra_cme} are not considered to interact with the CMEs from the main list in Table \ref{table_eventslist}, but occurred before them. The CME's main characteristics presented in the table were obtained through StereoCat fitting. 
\begin{table}[h]
\caption{List of CMEs that were not initially part of the study, but were added to the runs to create more realistic solar wind conditions. StereoCat was used to collect CMEs parameters.}
 \resizebox{\hsize}{!}{
\begin{tabular}{c|c|c|c|c|c|c}
\hline
 Main & Passage at 21.5 $R_{\text{Sun}}$&	Longitude	& Latitude	& Half Width  & Speed \\ Event No &
		& [$^\circ$]	& [$^\circ$]	&	[$^\circ$]		 & [km~s$^{-1}$]\\
\hline	
5 & 2012-03-05 06:58:00 & -28 & 45& 51&	1352 \\
\hline
6 & 2012-03-09 10:24:00 & 21 & 5.5 & 39 & 1049 \\
\hline
6 & 2012-03-10 20:32:00 & 11 & 14 & 44 & 1325 \\
\hline
14 & 2014-09-09 04:00:00 & -24 & 22 & 50 & 966 \\ \hline 
\end{tabular}}
\label{extra_cme}
\end{table}
\section{EUHFORIA runs - visualization of selected events}
\label{events_plots}
In the appendix, a series of plots is provided for each of selected event. These plots combine observational data obtained from OMNI, with the results of EUHFORIA simulations for both cone and spheromak models. Each line in the plots is colour-coded for clarity: purple lines represent cone simulation results, green lines depict spheromak simulation with a standard setup density of 1e-18. Additionally, red and blue lines correspond to spheromak simulation with an increased setup density of 1e-17 and 5e-18 respectively. Actual observational data is represented by black lines. To aid interpretation, vertical lines of corresponding colours indicate the arrival time of the ICME within each plot.
The $K_p$ value for all scenarios is determined using the Eq. \ref{Kpeq}, by \cite{2008Newell}. 

\begin{figure}[ht]
\centering
\resizebox{\hsize}{!}{
\includegraphics{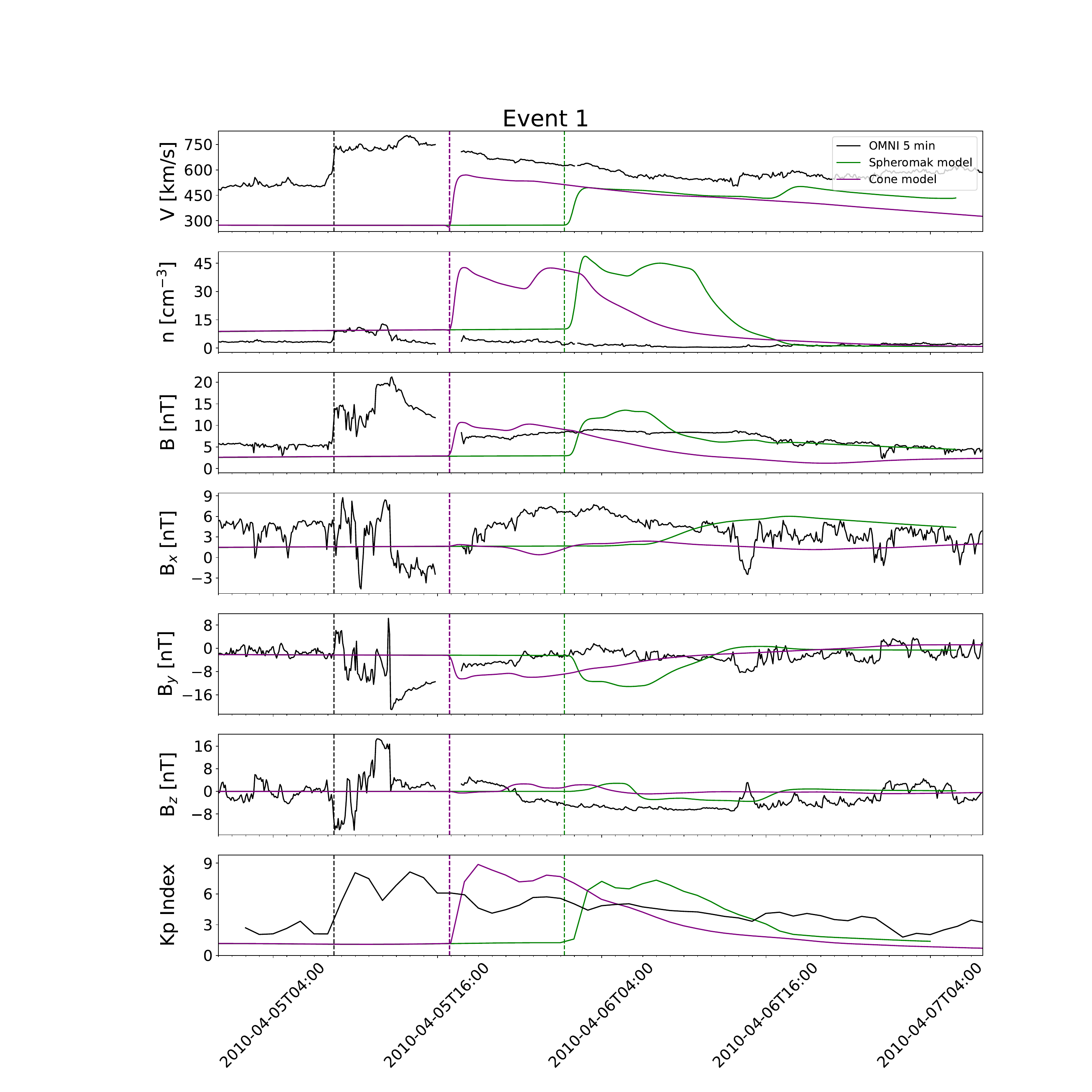}}
\caption{Comparison of real data with EUHFORIA cone and spheromak runs for Event 1.} \label{Event1}
\end{figure}

\begin{figure}[ht]
\centering
\resizebox{\hsize}{!}{\includegraphics{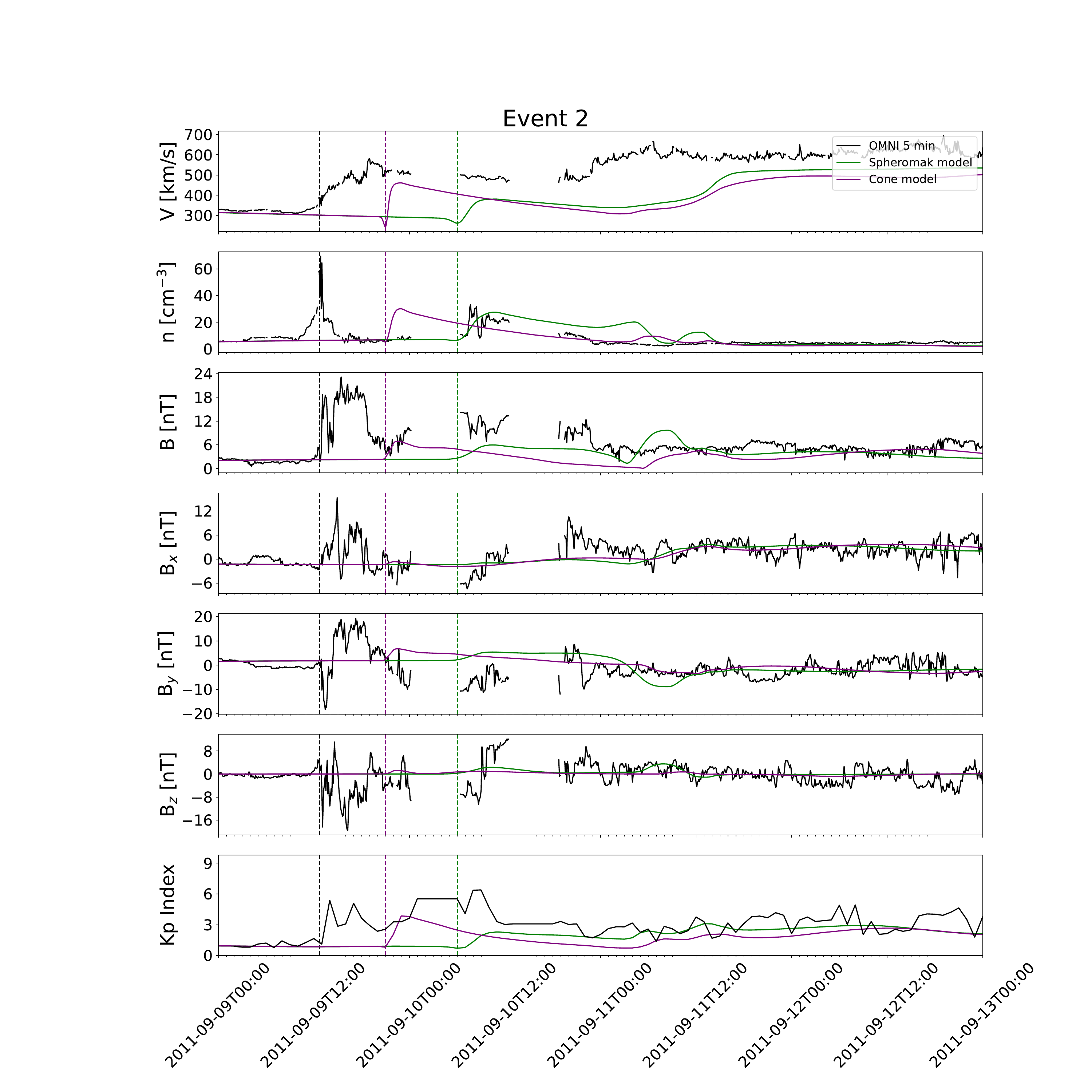}}
\caption{Comparison of real data with EUHFORIA cone and spheromak runs for Event 2} \label{Event2}
\end{figure}

\begin{figure}[ht]
\centering
\resizebox{\hsize}{!}{\includegraphics{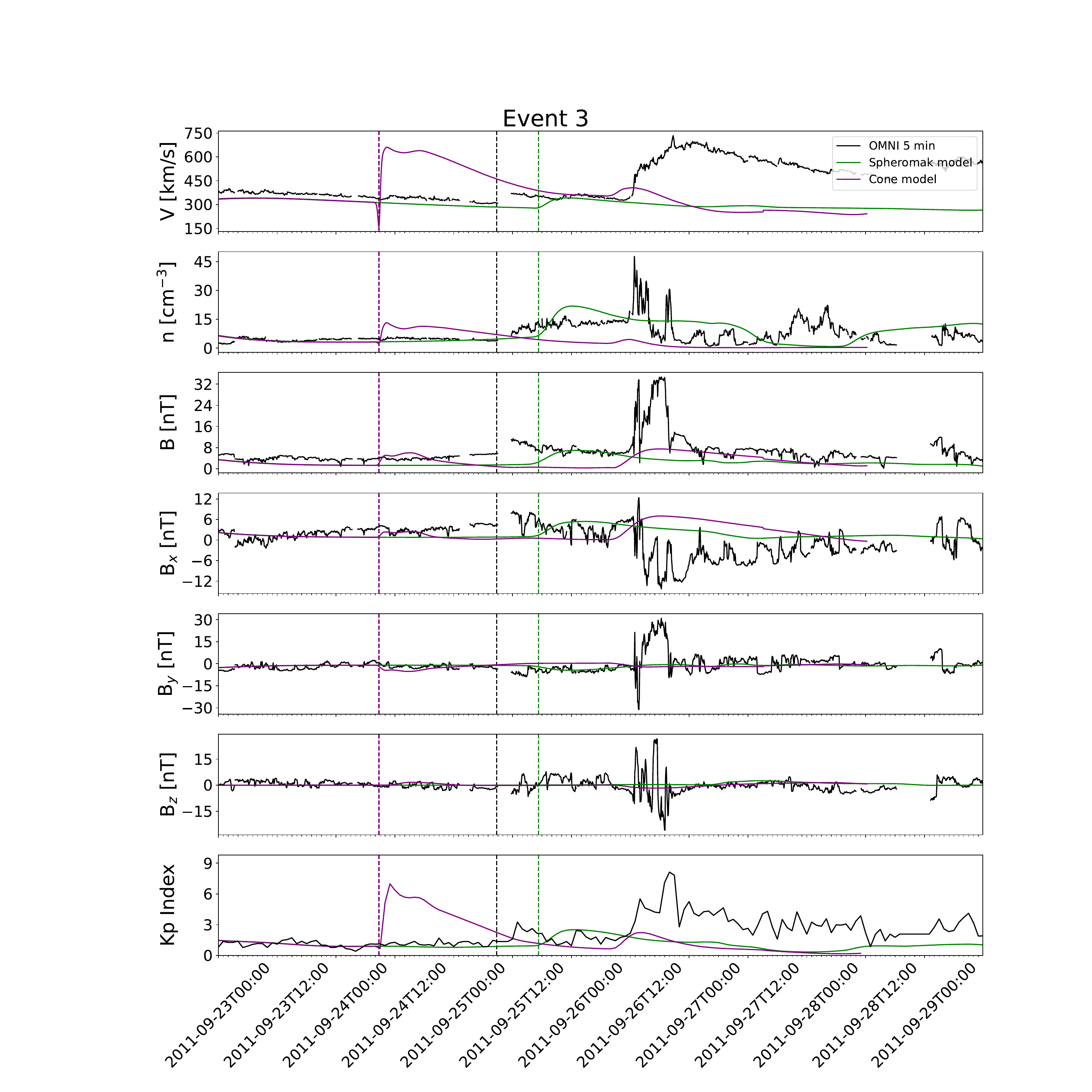}}
\caption{Comparison of real data with EUHFORIA cone and spheromak runs for Event 3} \label{Event3}
\end{figure}
\begin{figure}[ht]
\centering
\resizebox{\hsize}{!}{\includegraphics{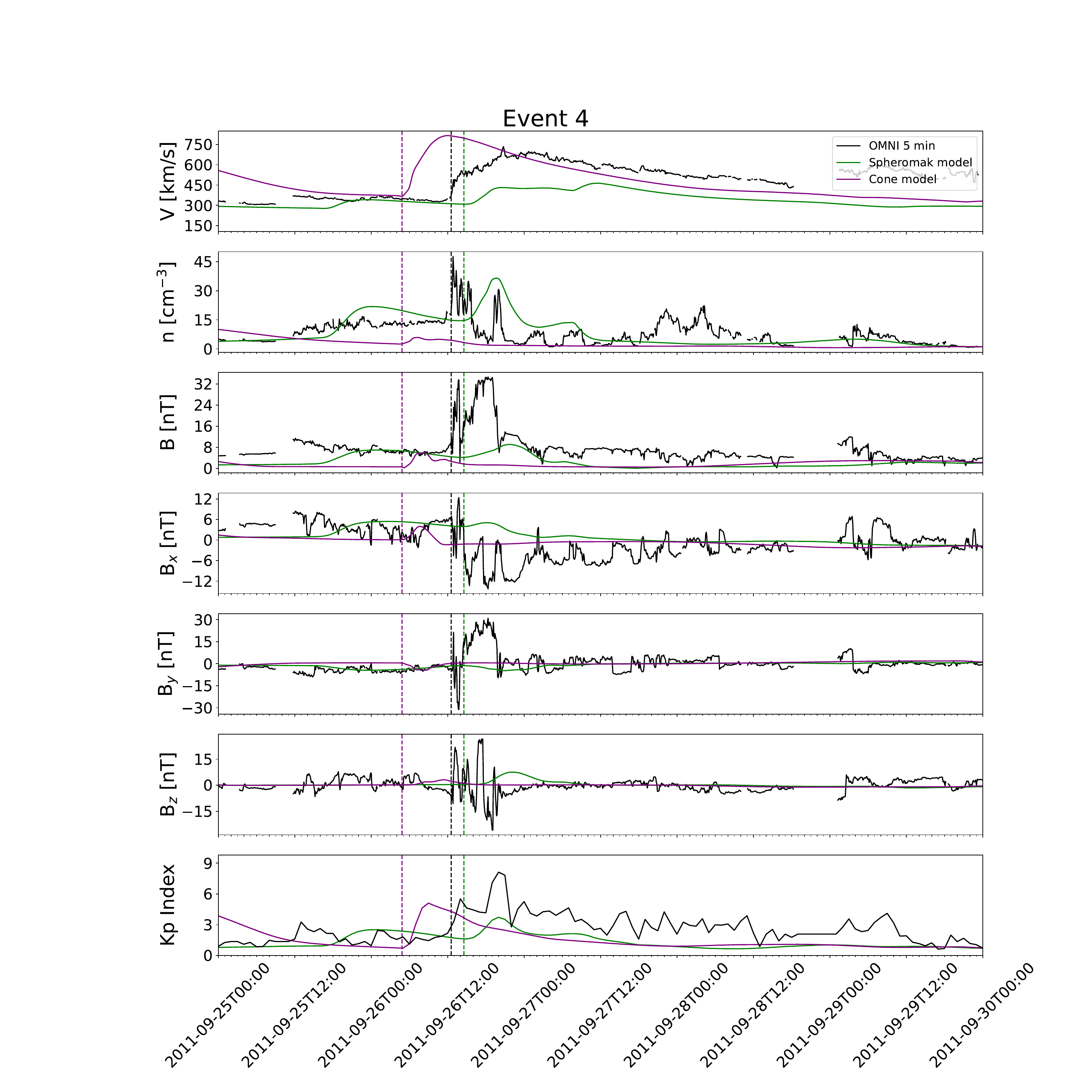}}
\caption{Comparison of real data with EUHFORIA cone and spheromak runs for Event 4} \label{Event4}
\end{figure}
\begin{figure}[ht]
\centering
\resizebox{\hsize}{!}{\includegraphics{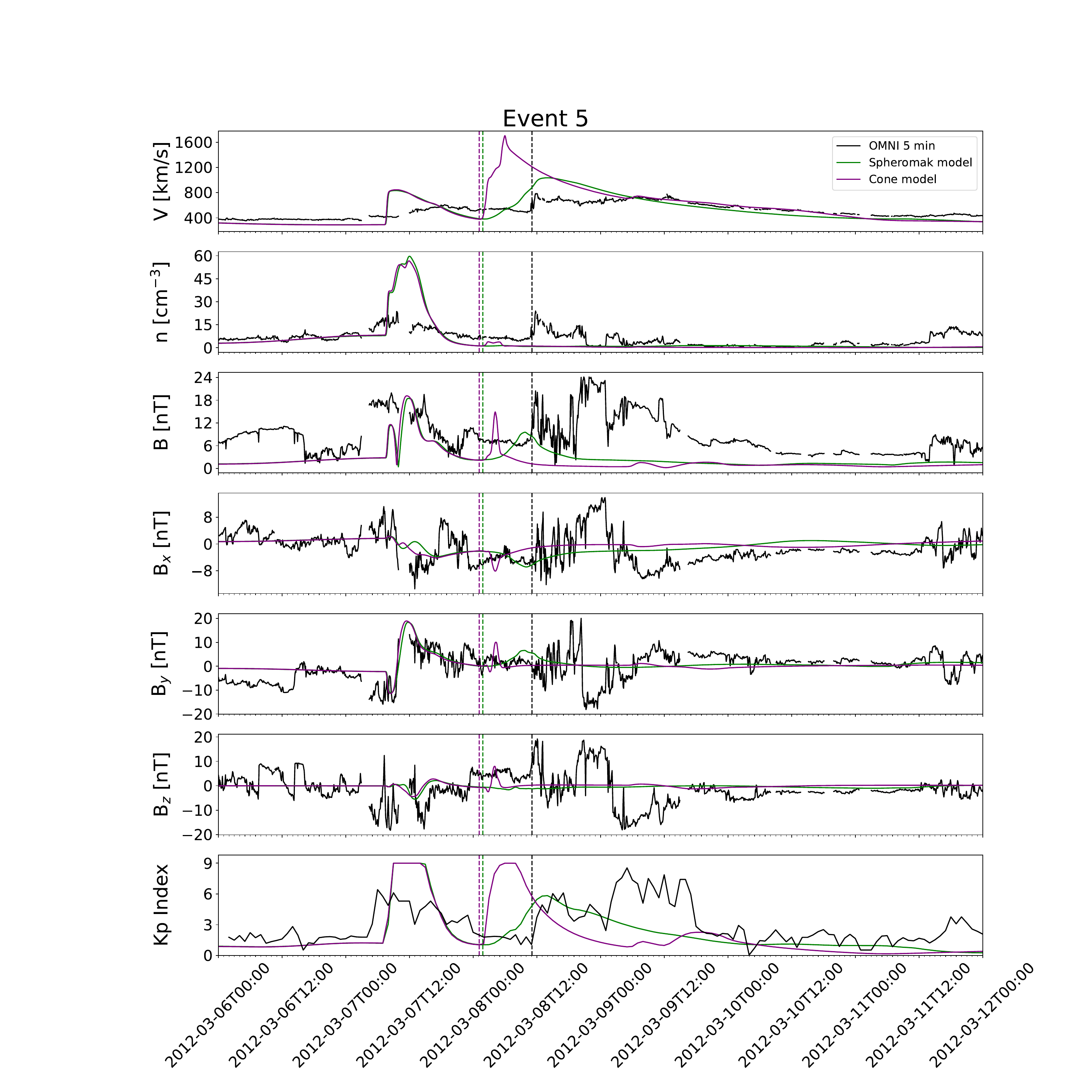}}
\caption{Comparison of real data with EUHFORIA cone and spheromak runs for Event 5} \label{Event5}
\end{figure}
\begin{figure}[ht]
\centering
\resizebox{\hsize}{!}{\includegraphics{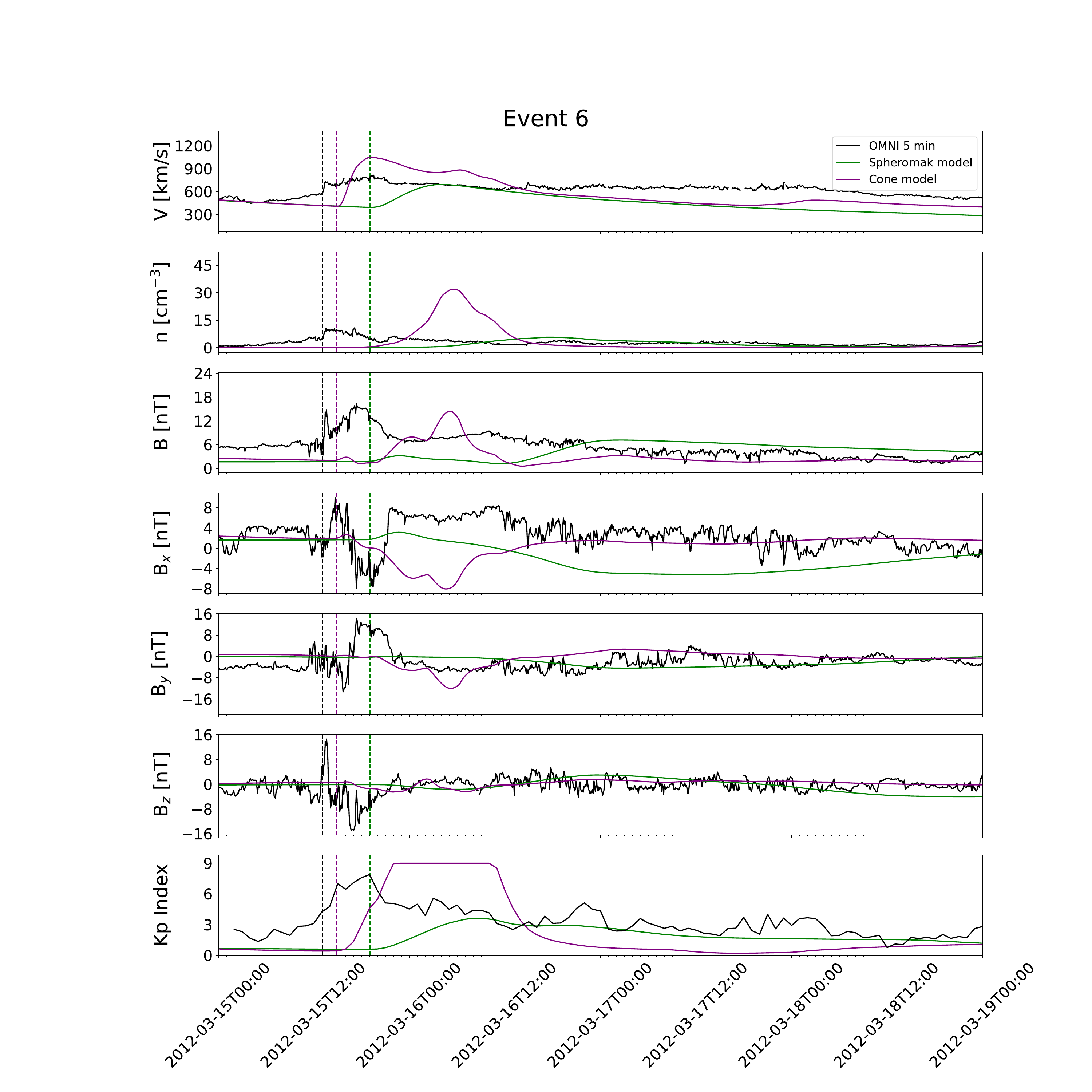}}
\caption{Comparison of real data with EUHFORIA cone and spheromak runs for Event 6} \label{Event6}
\end{figure}
\begin{figure}[ht]
\centering
\resizebox{\hsize}{!}{\includegraphics{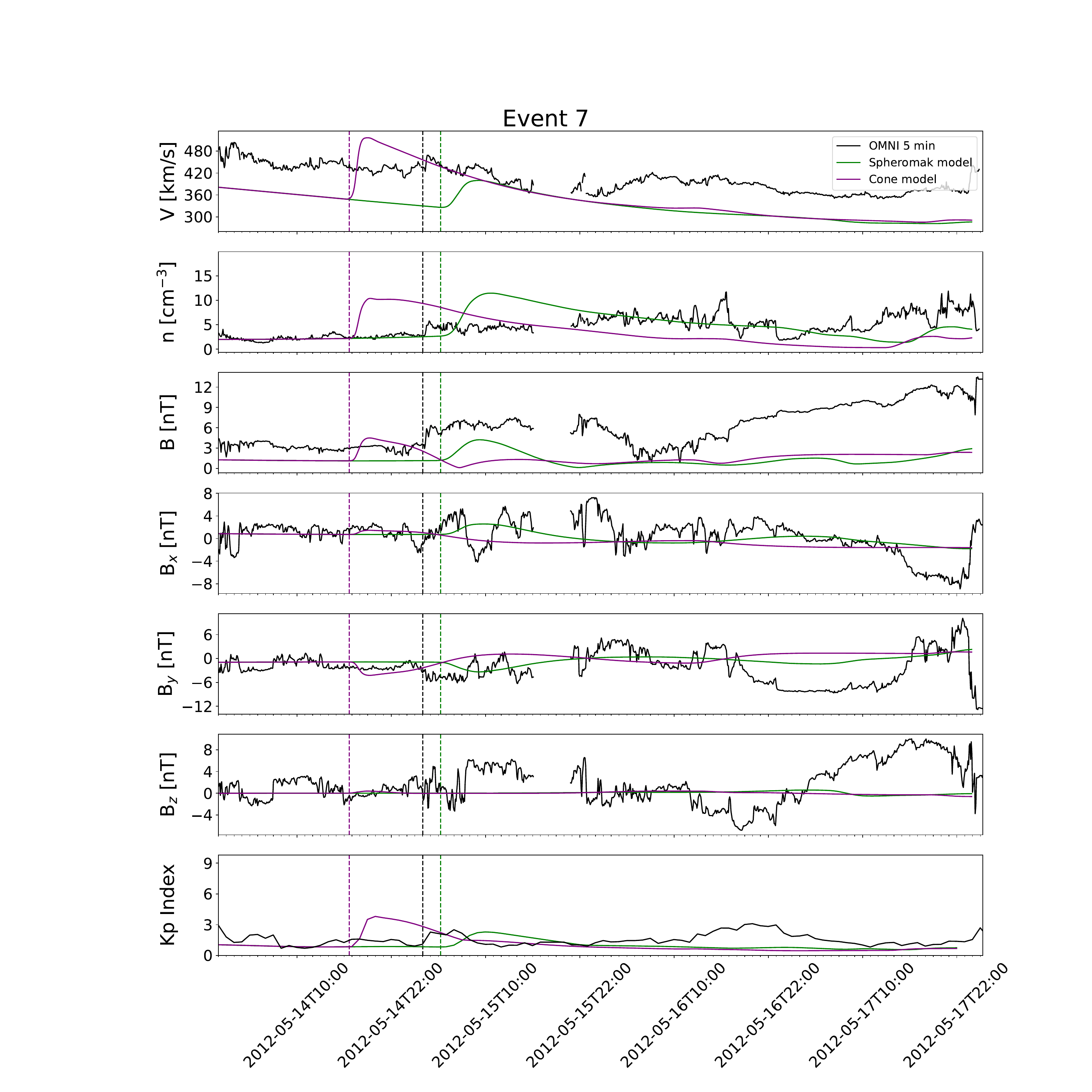}}
\caption{Comparison of real data with EUHFORIA cone and spheromak runs for Event 7} \label{Event7}
\end{figure}
\begin{figure}[ht]
\centering
\resizebox{\hsize}{!}{\includegraphics{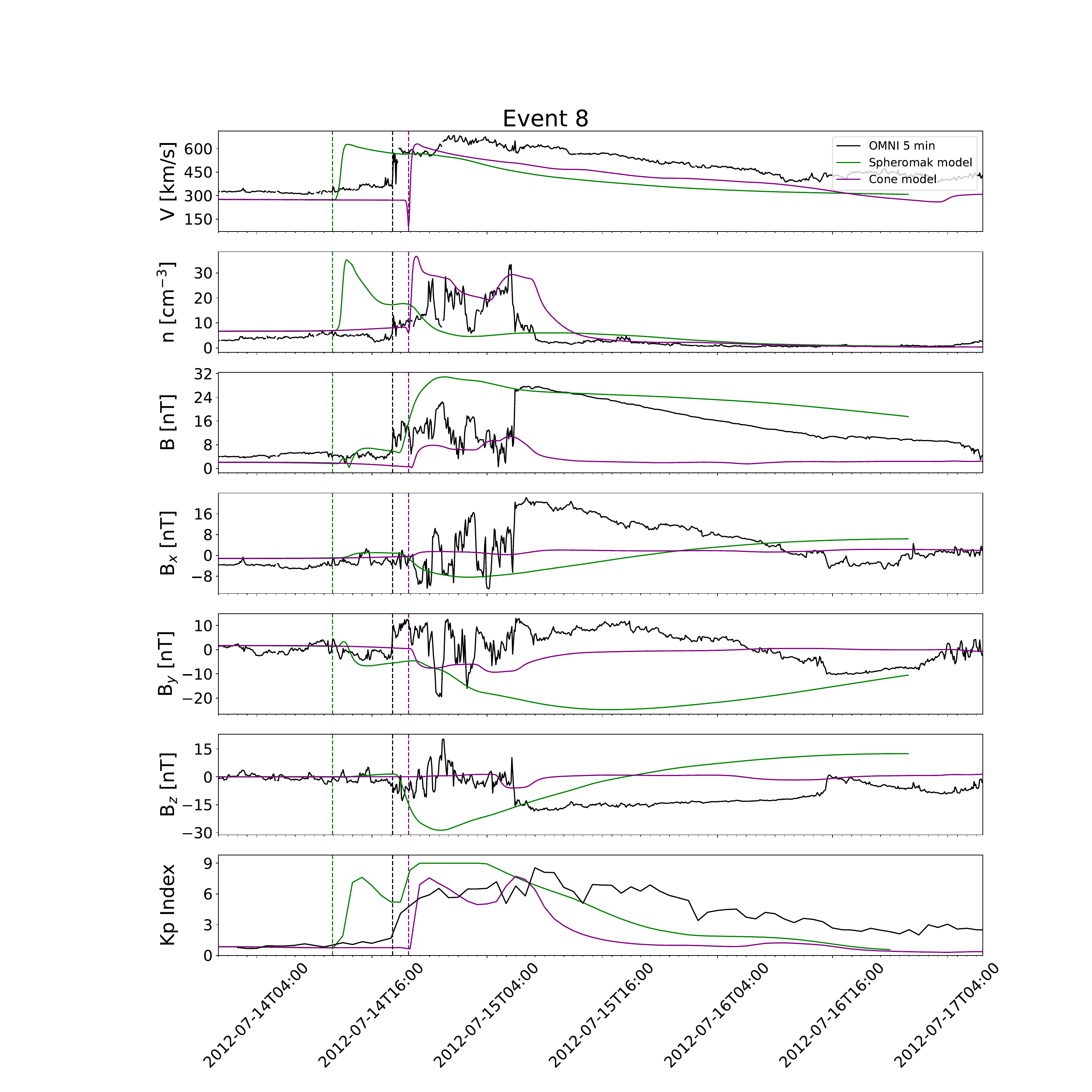}}
\caption{Comparison of real data with EUHFORIA cone and spheromak runs for Event 8} \label{Event8}
\end{figure}
\begin{figure}[ht]
\centering
\resizebox{\hsize}{!}{\includegraphics{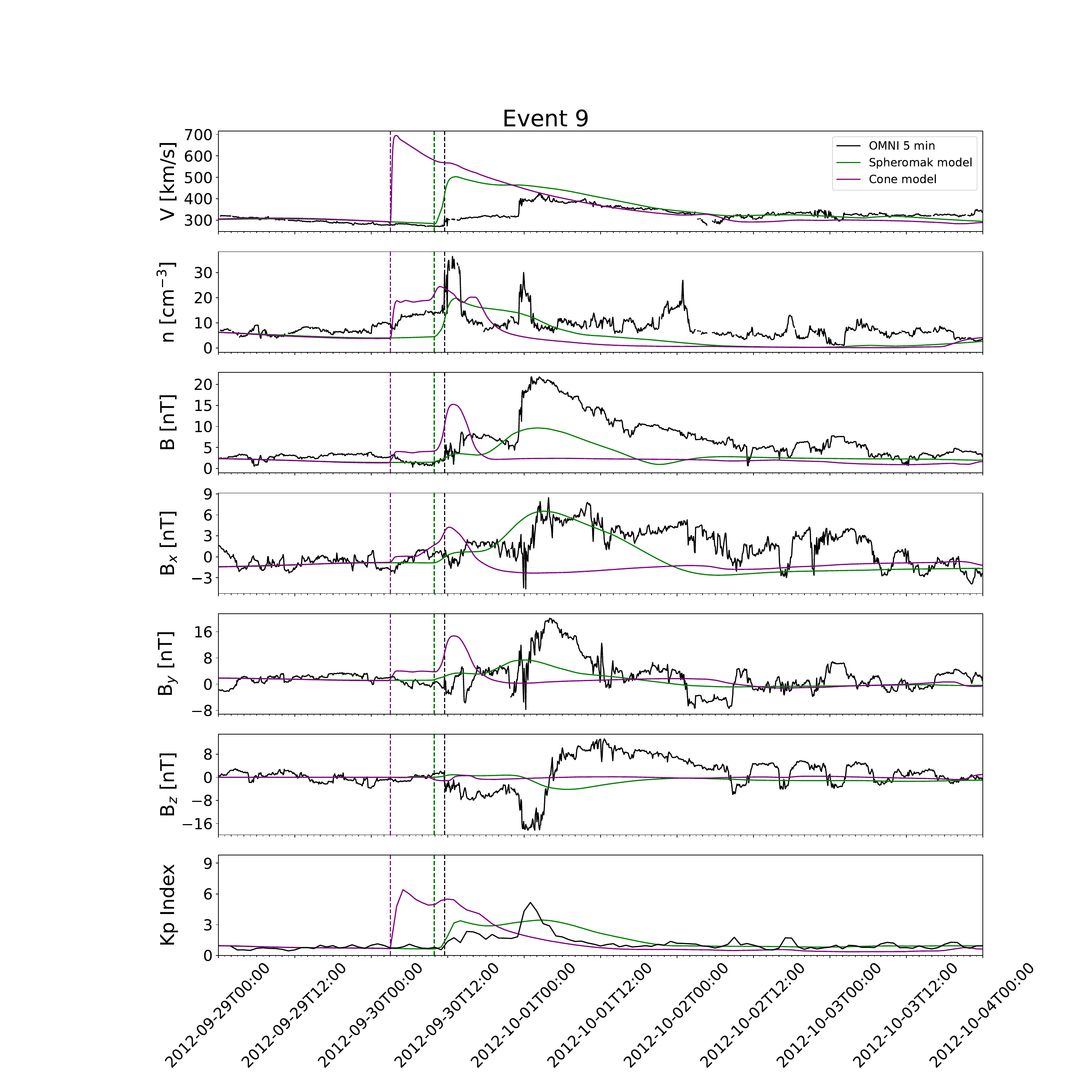}}
\caption{Comparison of real data with EUHFORIA cone and spheromak runs for Event 9} \label{Event9}
\end{figure}
\begin{figure}[ht]
\centering
\resizebox{\hsize}{!}{\includegraphics{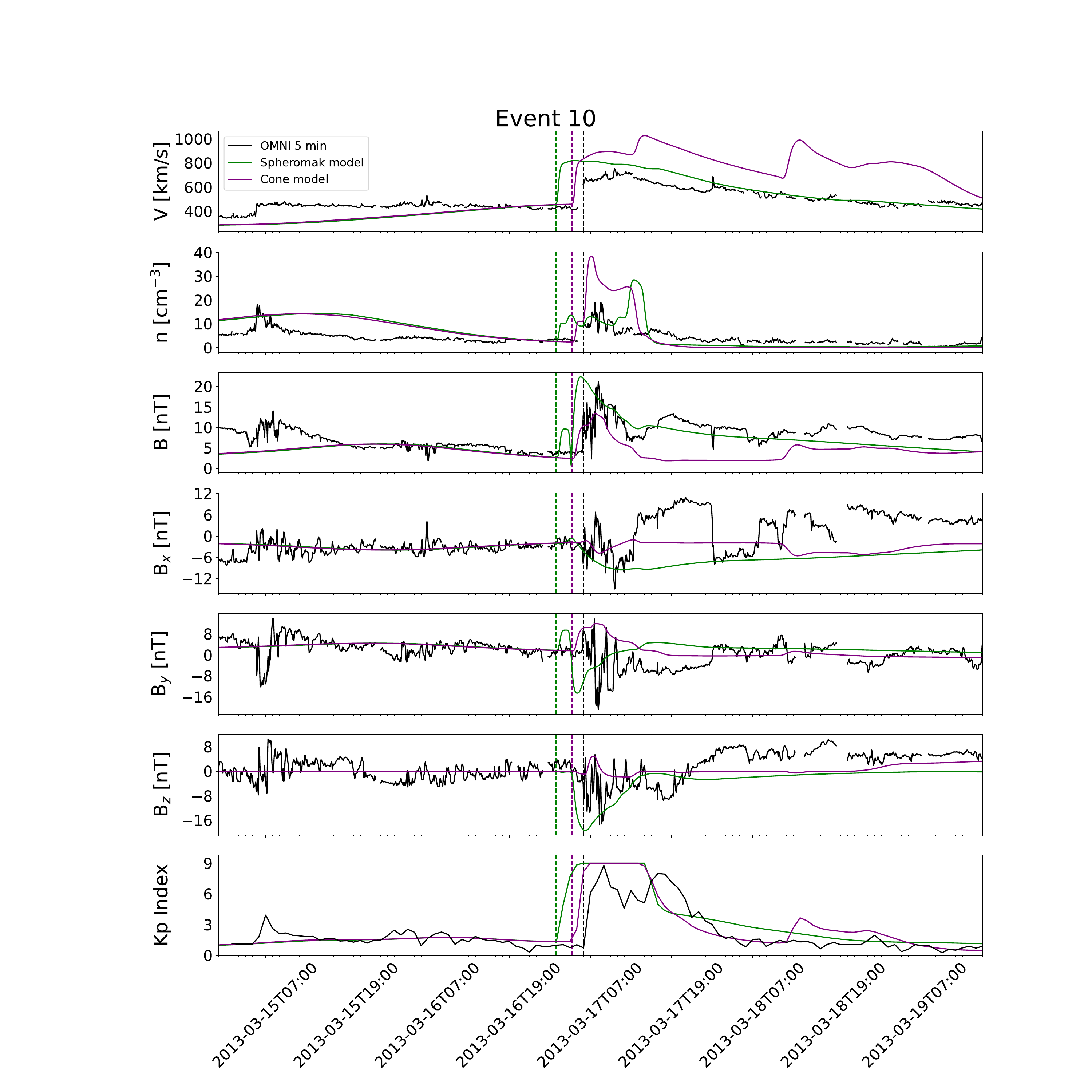}}
\caption{Comparison of real data with EUHFORIA cone and spheromak runs for Event 10} \label{Event10}
\end{figure}
\begin{figure}[ht]
\centering
\resizebox{\hsize}{!}{\includegraphics{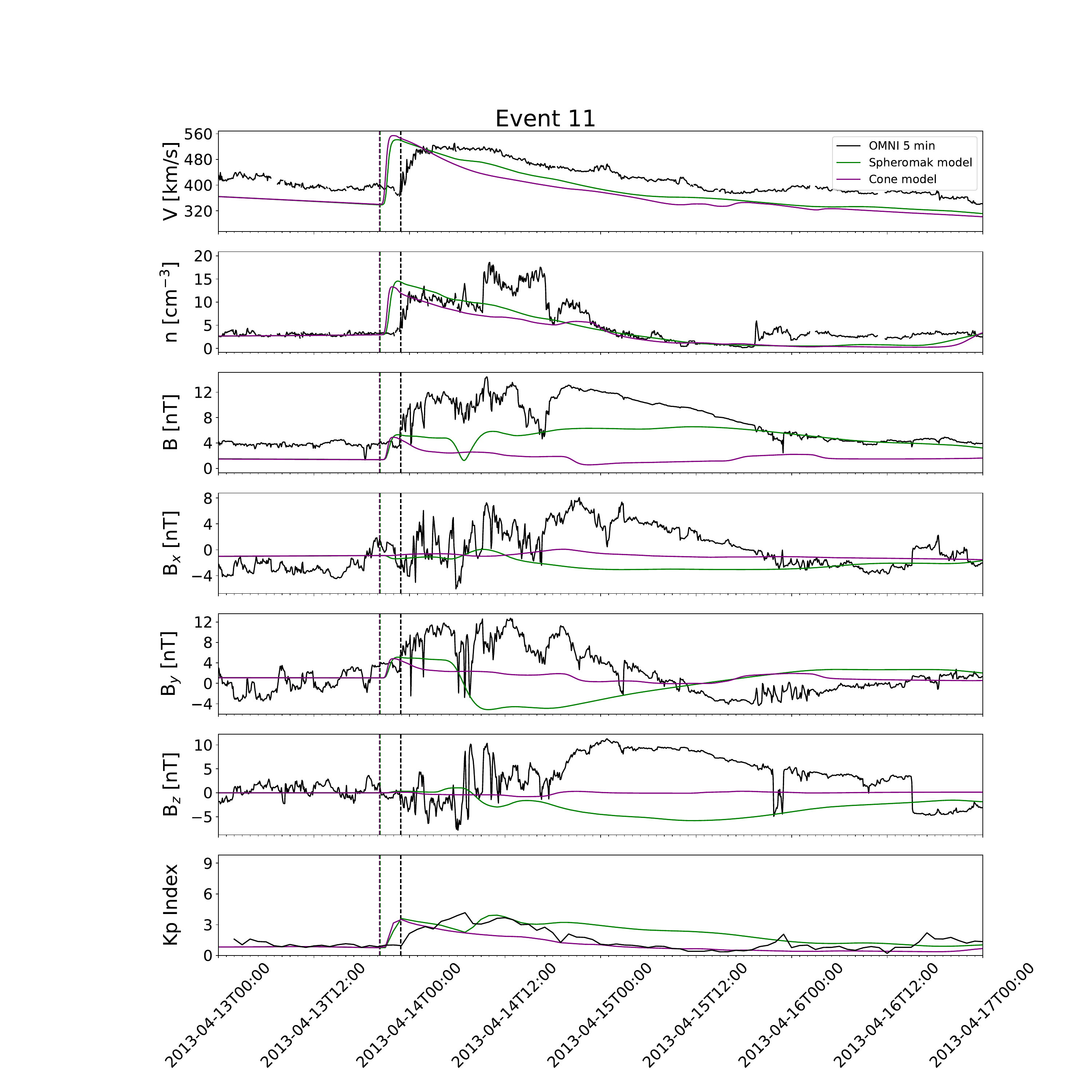}}
\caption{Comparison of real data with EUHFORIA cone and spheromak runs for Event 11} \label{Event11}
\end{figure}
\begin{figure}[ht]
\centering
\resizebox{\hsize}{!}{\includegraphics{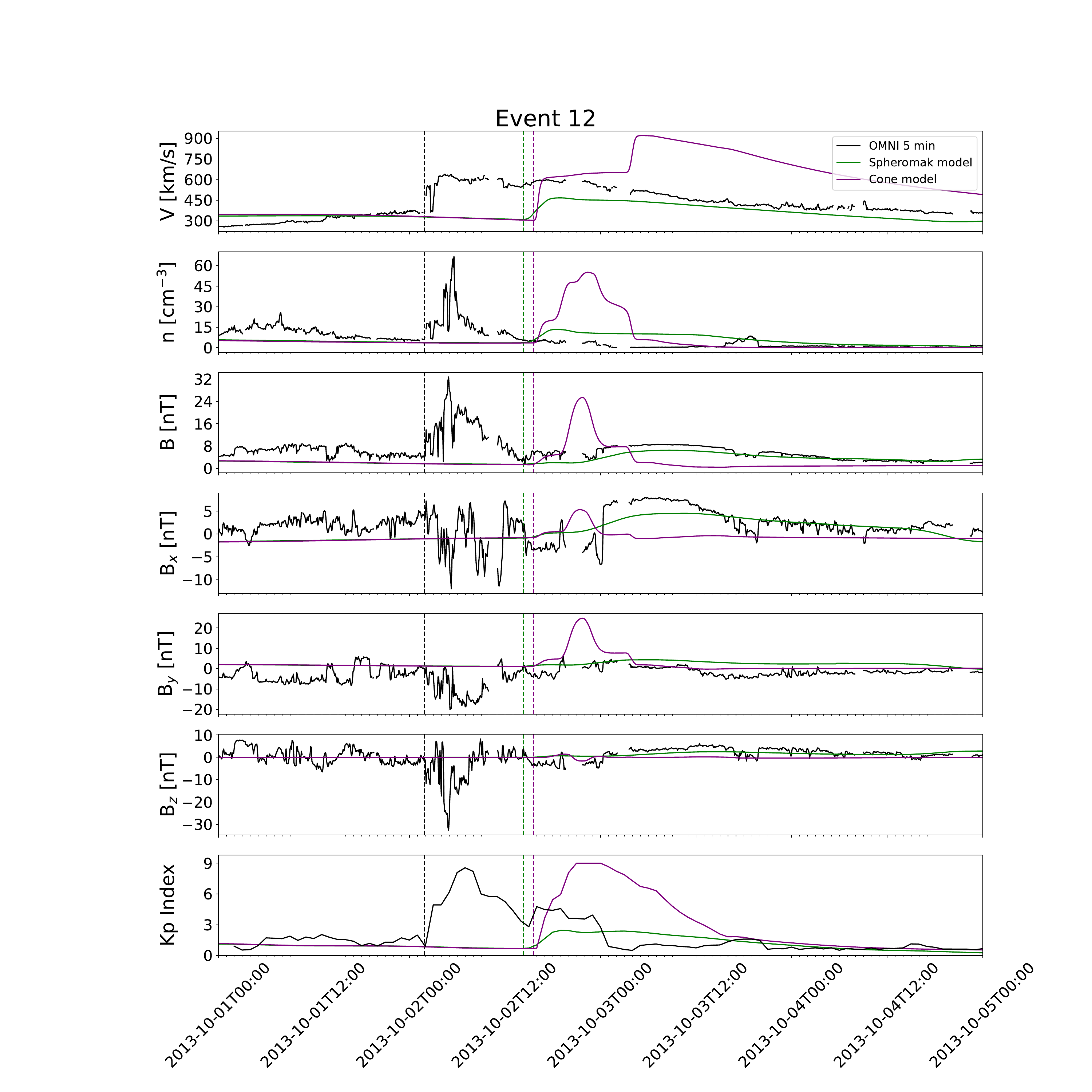}}
\caption{Comparison of real data with EUHFORIA cone and spheromak runs for Event 12} \label{Event12}
\end{figure}
\begin{figure}[ht]
\centering
\resizebox{\hsize}{!}{\includegraphics{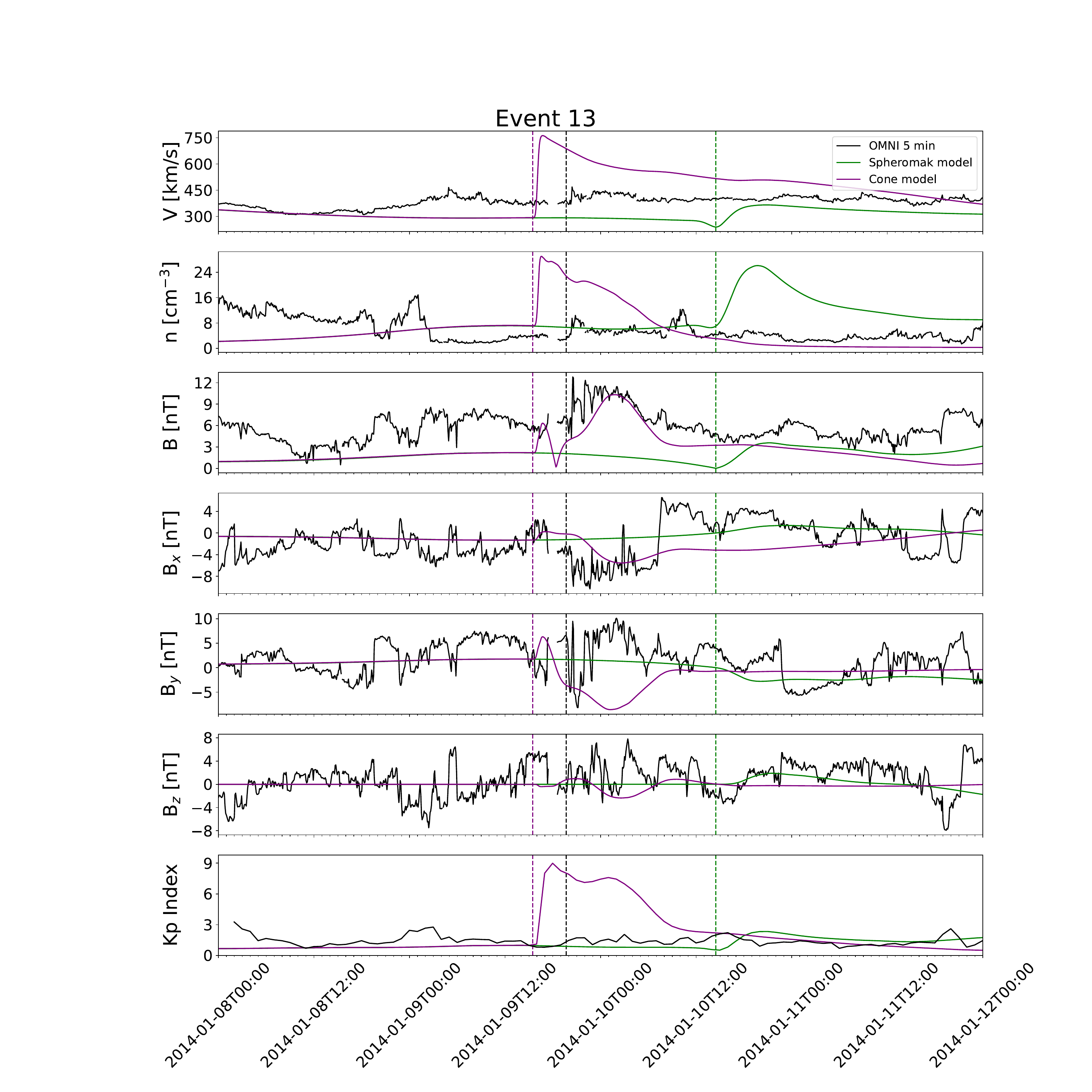}}
\caption{Comparison of real data with EUHFORIA cone and spheromak runs for Event 13} \label{Event13}
\end{figure}
\begin{figure}[ht]
\centering
\resizebox{\hsize}{!}{\includegraphics{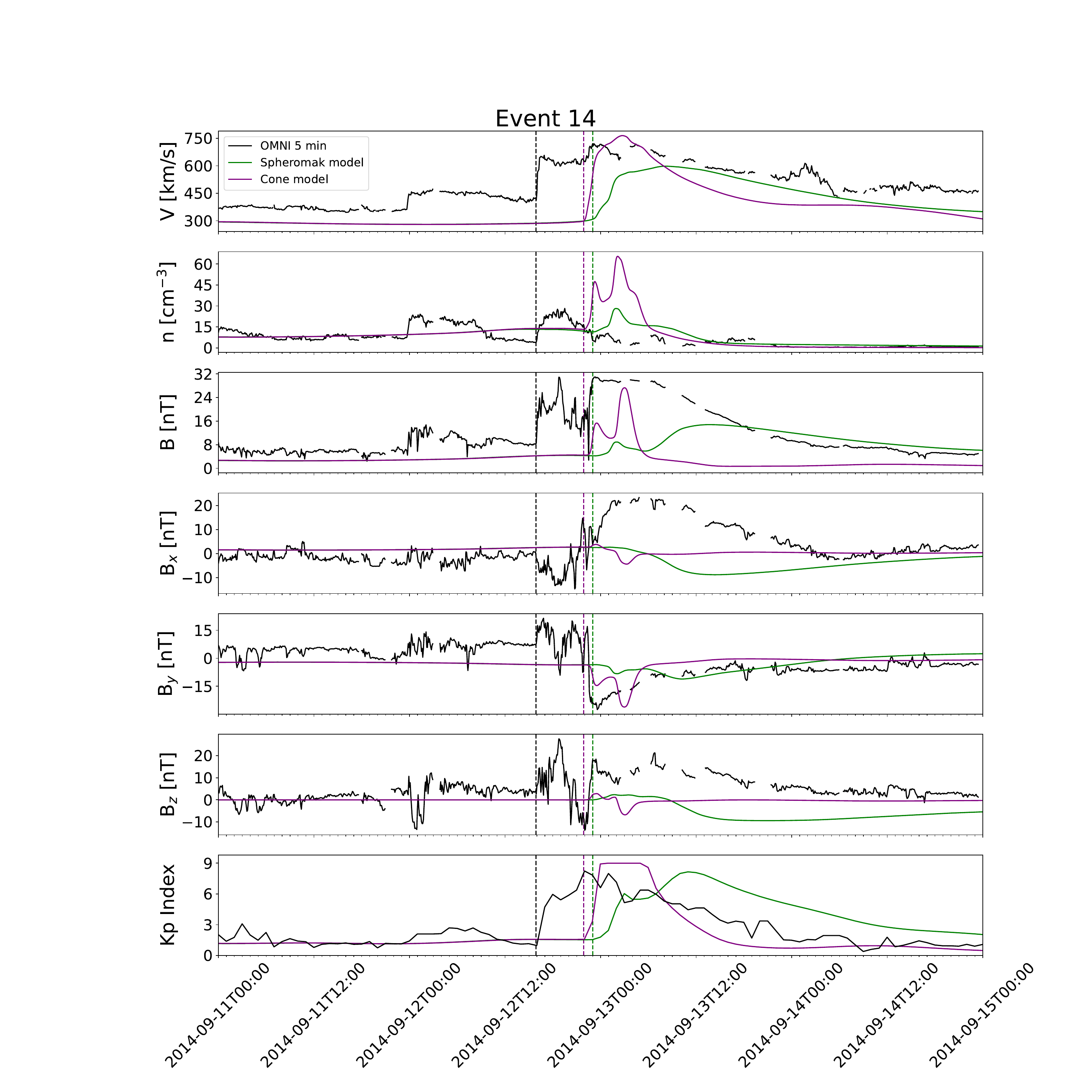}}
\caption{Comparison of real data with EUHFORIA cone and spheromak runs for Event 14} \label{Event14}
\end{figure}
\begin{figure}[ht]
\centering
\resizebox{\hsize}{!}{\includegraphics{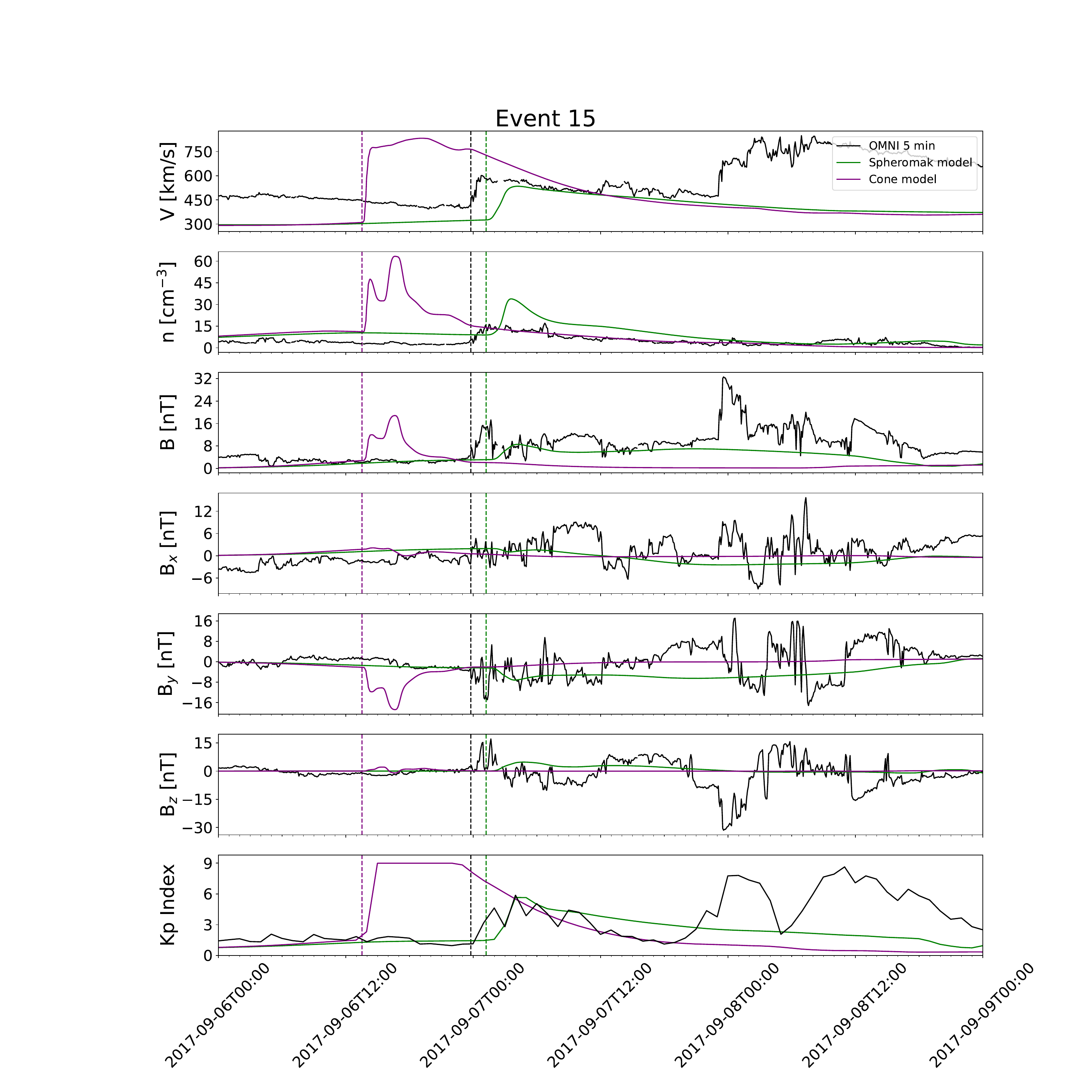}}
\caption{Comparison of real data with EUHFORIA cone and spheromak runs for Event 15} \label{Event15}
\end{figure}
\begin{figure}[ht]
\centering
\resizebox{\hsize}{!}{\includegraphics{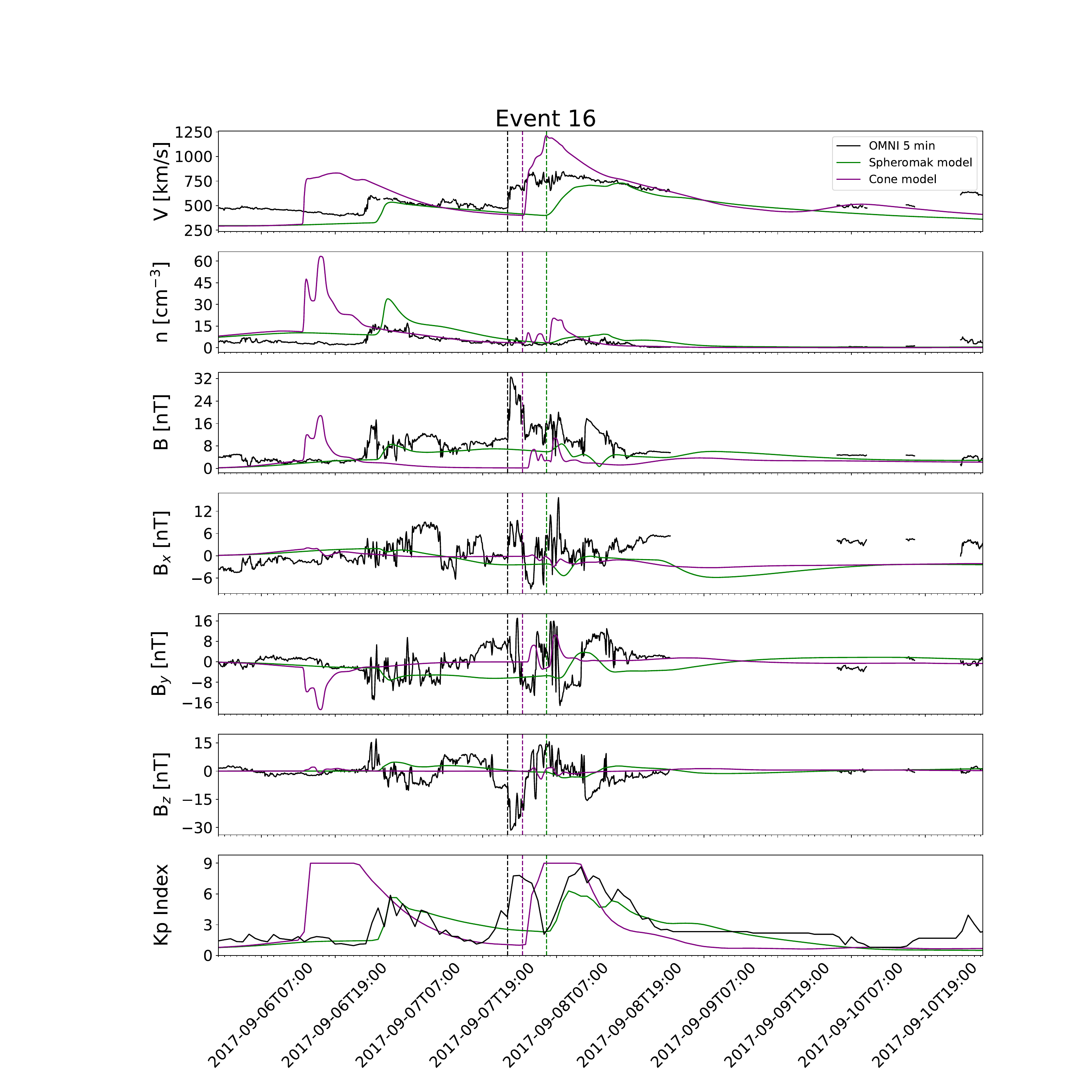}}
\caption{Comparison of real data with EUHFORIA cone and spheromak runs for Event 16} \label{Event16}
\end{figure}

\end{appendix}

\end{document}